\documentstyle[seceq,preprint,epsf]{jpsj}

\def\runtitle{Theory of Pseudogap Phenomena in High-$T_{c}$ cuprates 
based on the strong coupling superconductivity}
\def\runauthor{Youichi {\sc Yanase}, Kosaku {\sc Yamada} }

\title{ Theory of Pseudogap Phenomena in High-$T_{{\rm c}}$ Cuprates \\
Based on the Strong Coupling Superconductivity}

\author{Youichi {\sc Yanase}\footnote{E-mail: yanase@ton.scphys.kyoto-u.ac.jp}
 and Kosaku {\sc Yamada}}

\inst{Department of Physics, kyoto University, Kyoto 606-8502}

\recdate{April 12, 1999}

\abst
{
 In this paper we study the effects of the strong coupling superconductivity 
on the normal state electronic structure. We point out that the pseudogap 
phenomena in High-$T_{{\rm c}}$ cuprates are naturally explained 
as a precursor of the strong coupling superconductivity. 
 We expand the reciprocal of the T-matrix 
with respect to the momentum and the frequency, 
and we discuss the properties of the expansion 
parameters. We estimate these parameters and carry out 
the calculations for the single particle self-energy 
using the T-matrix and 
the self-consistent T-matrix approximations, respectively. 
 We show that the superconducting fluctuations, which are strongly diffusive 
within the conventional weak coupling theory, become propagative and have 
the character of the pre-formed pairs in the strong coupling case. 
 The spectral weight and the density of states show the gap-like features 
by the effects of the resonances with the thermally excited weakly-damped 
pre-formed pairs. 
 The spectral weight at the Fermi energy is strongly suppressed. 
 This suppression starts at the mean field critical 
temperature $T_{{\rm MF}}$ and the actual critical temperature 
$T_{{\rm c}}$ is strongly reduced owing to the fluctuations. 
 These features properly explain to the pseudogap phenomena 
in High-$T_{{\rm c}}$ cuprates. 
}

\kword
{high-$T_{{\rm c}}$ cuprates, pseudogap phenomena, 
strong coupling superconductivity, \\
superconducting fluctuation, weakly damped pre-formed pair, resonance} 

\begin{document}
\sloppy
\maketitle

\section{Introduction}

 Since the discovery of high-temperature (High-$T_{{\rm c}}$) 
superconductivity by Bednortz and M$\rm{\ddot{u}}$ller,~\cite{rf:bednortz} 
the anomalous normal state properties have been studied 
from the various points of view. 

 In particular, the pseudogap phenomena in under-doped cuprates have been 
a very important issue. There are a lot of studies for the issue from both  
experimental and theoretical points of view. However, 
the complete understanding remains to be obtained. 

 The pseudogap phenomena mean the suppression of the low frequency spectral 
weight without any long range order. They are universal phenomena observed in 
various compounds of under-doped cuprates. 

 The normal state excitation gap in under-doped cuprates have been indicated 
by several various experiments. 
 The nuclear magnetic resonance (NMR) experiments have shown the anomalous 
temperature dependence of the spin lattice relaxation rate $1/T_{1}$ and 
the spin susceptibility $\chi$.~\cite{rf:NMR,rf:isida} 
 The quantity $1/T_{1}T$, which increases with decreasing temperature, 
starts to decrease at the pseudogap onset temperature $T^{*}$.~\cite{rf:NMR} 
 The spin susceptibility decreases gradually from the rather high 
temperature, and immediately changes its slope at $T^{*}$.~\cite{rf:isida} 
 
 The optical conductivity have indicated the suppression of the low frequency 
spectral weight above the superconducting critical temperature 
$T_{{\rm c}}$.~\cite{rf:homes} The suppression is continuous above and below 
$T_{{\rm c}}$. 

 The transport coefficients also change their behaviors at $T^{*}$. 
 The T-linear in-plane resistivity observed in under-doped cuprates deviates 
downward.~\cite{rf:ito,rf:oda} The Hall coefficient remarkably deviates 
downward and decreases with temperature.~\cite{rf:ito} The c-axis resistivity 
strongly increases in the pseudogap region.~\cite{rf:takenaka} 
 It have been shown that these behaviors of the transport phenomena are 
naturally explained by considering the momentum dependent scattering rate 
owing to the anti-ferromagnetic 
spin fluctuations.~\cite{rf:stojkovic,rf:yanase}

 In particular, the angle-resolved photo-emission spectrum 
(ARPES)~\cite{rf:ARPES} experiments have 
directly shown the suppression of the low frequency spectral weight. 
 Moreover, ARPES experiments have shown that the shape of the pseudogap 
is similar to that of the superconducting gap, 
and the magnitude does not change 
at the superconducting critical point.~\cite{rf:norman} 
 After all, the pseudogap has the $d$-wave shape and is continuously 
connected with the superconducting gap. 

 It is shown that the onset temperature $T^{*}$ measured by the 
above experiments is almost the same as the mean field superconducting 
critical temperature $T_{{\rm MF}}$ estimated from 
the amplitude of the superconducting gap.~\cite{rf:oda}

 The recent tunneling experiments have directly shown the suppression of 
the low frequency density of states.~\cite{rf:renner}  
 In this connection, the gap-like structure which is similar to 
the normal state pseudogap is observed in vortex cores in the 
superconducting states under the high magnetic fields.~\cite{rf:renner2} 

 The impurity effects on the reduction of $T_{{\rm c}}$ have indicated the 
suppression of the low frequency density of states.~\cite{rf:tallon} 

 The electronic specific heat is reduced well above $T_{{\rm c}}$ and $T^{*}$ 
and the step height at $T_{{\rm c}}$ is extraordinary small 
in under-doped cuprates.~\cite{rf:loram} 
 This means that the entropy has been already lost at rather high temperature.

 From the above various experiments suggesting the relevance and continuity 
to the superconductivity, we consider that the pseudogap phenomena are the 
precursor of the strong coupling superconductivity. 
 In this paper, we explain the pseudogap phenomena 
on the basis of the strong coupling superconductivity.

 Several theoretical proposals have been given for the pseudogap phenomena. 
 Here, we give a brief review on some important proposals. 

 One of them is the resonating valence bond (RVB) theory.~\cite{rf:tanamoto} 
 This theory is based on the non-Fermi liquid state in which the distinct 
excitations, spinon and holon, exist. 
 In the RVB theory, the spinon pairing occurs at $T^{*}$ 
(so-called 'spin gap'), and 
holons condense at $T_{{\rm c}}$. Since 'spin gap' is based on the 
singlet pairing, the RVB theory explains the various experiments for 
under-doped cuprates. 
 However, the continuities from the normal states to the superconducting 
states and from over-doped to under-doped cuprates are 
not necessarily obvious, 
and the physical origin of the spin-charge separation is not clear. 
 
 The magnetic scenarios based on the anti-ferromagnetic or SDW gap formation 
or their precursor have been proposed 
by various authors.~\cite{rf:SDW,rf:dahmSDW} 
 In these theories, 'hot spot', which is a part of the Fermi surface 
in the vicinity of the anti-ferromagnetic Brillouin zone, 
exists near $(0,\pi)$ and plays an especial role. 
 Since the quasiparticles at 'hot spot' are strongly scattered 
by the anti-ferromagnetic spin fluctuations, 
the gap structure appears at 'hot spot'. 
 However, considering that the ground state is actually superconductive, 
we can hardly expect that the magnetic order and 
the superconductivity are continuously connected 
beyond the superconducting transition. 
 We shall give a discussion about this point in the last section. 

 Next, we describe the pairing scenarios based on the strong coupling 
superconductivity. 
 Generally, the strong coupling superconductivity indicates the existence of 
the incoherent Cooper pairs (pre-formed pairs), 
as the famous Nozi$\grave{{\rm e}}$res and Schmitt-Rink 
formalism~\cite{rf:Nozieres} has described. 
 The strong attractive interaction produces the pre-formed pairs above the 
superconducting critical point. The pre-formed pairs condense at the 
critical point. 

 Our scenario in this paper is different from such a simple viewpoint. 
 We think of the pseudogap as the phenomena brought about 
by the strong superconducting fluctuations. We actually carry out the 
different calculations from the Nozi$\grave{{\rm e}}$res and Schmitt-Rink 
formalism. 
 The strong coupling superconductivity naturally gives rise to 
the strong fluctuations and the deviations from the BCS mean field 
descriptions. Because the short coherence length, which is attributed to 
the high critical temperature $T_{{\rm c}}$ compared with the renormalized 
Fermi energy $\tilde{\varepsilon}_{{\rm F}}$, 
and the quasi two-dimensionality characterize the superconductivity of 
High-$T_{{\rm c}}$ cuprates, it is natural to consider the 
strong superconducting fluctuations. 

 The strong coupling superconductivity has been discussed from the view point 
of the crossover from the BCS superconductivity to the 
Bose-Einstein condensation both in the ground state~\cite{rf:leggett} and 
at the finite temperature.~\cite{rf:Nozieres} 
 The concept of the Nozi$\grave{{\rm e}}$res and Schmitt-Rink 
formalism~\cite{rf:Nozieres} is taking account of the corrections to the 
thermodynamic potential shown in Fig.1 and deciding the chemical potential 
self-consistently. 
 This procedures correspond to counting the number of the bosonic pairs. 
 Their formalism is justified in the low density limit. 
 Several authors have discussed the applicability of 
the Nozi$\grave{{\rm e}}$res and Schmitt-Rink formalism 
to the two-dimensional systems.~\cite{rf:schmitt-rink,rf:tokumitu} 

\begin{figure}[ht]
  \begin{center}
%   \figureheight{1.0cm}  
   \epsfysize=4cm
$$\epsffile{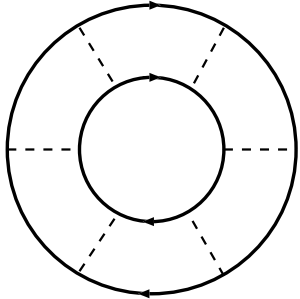}$$
%    \epsfile{file=Fig1,height=4cm}
    \caption{The correction to the thermodynamic potential introduced in the 
             Nozi$\grave{{\rm e}}$res and Schmitt-Rink theory.}
    \label{fig:Nozieres and Schmitt-rink theory}
  \end{center}
\end{figure}

 The precursor of the superconductivity has been proposed for 
High-$T_{{\rm c}}$ cuprates by Randeria {\it et al.}~\cite{rf:randeriareview} 
on the basis of the functional integral formulation for $s$-wave symmetry. 
 The Ginzburg-Landau theory is derived for $d$-wave symmetry by Stintzing and 
Zwerger.~\cite{rf:stintzing}
 The self-consistent calculation has been carried out 
by Haussmann.~\cite{rf:haussmann} 
 Their theories have been based on the 
Nozi$\grave{{\rm e}}$res and Schmitt-Rink 
formalism~\cite{rf:Nozieres} and treated the Fermi gas model or 
the low density model. 
 The Nozi$\grave{{\rm e}}$res and Schmitt-Rink formalism has been applied to 
the $d-p$ model by Koikegami and Yamada.~\cite{rf:koikegami} 

 However, if anything, the nearly half-filled lattice system should be 
regarded as the rather high density system. 
 Moreover, since the pseudogap phenomena take place near the Fermi surface, 
the situation in which the Fermi surface remarkably changes and disappears 
at last is not realistic. 
 Therefore, we must take account of the higher order corrections in order to 
consider the effects of the superconducting fluctuations on the 
normal state electronic structure. 

 The phase fluctuation scenarios have been proposed by Emery and Kivelson,
~\cite{rf:emery} and calculated by other authors.~\cite{rf:kwon}
 The phenomenology based on the electron-like pre-formed pairs coexisting 
with unpaired fermions have been described 
by Geshkenbein {\it et al.}~\cite{rf:geshkenbein} 
with reference to the sign problem of the fluctuational Hall effect. 
 We shall shortly discuss the problem in the next section. 

 The explicit self-consistent T-matrix calculations for the electronic 
structure has been carried out numerically by several authors. 
~\cite{rf:engelbrecht,rf:ichinomiya} 
 However, their calculations deal with somewhat unrealistic situations 
such as the low density, the particle-hole symmetry and so on. 
 We shall point out the importance of their realistic factors later. 
 
 Maly {\it et al.}\cite{rf:maly} have introduced the idea of 'resonances' 
which means the scattering mechanism by the presence of the meta-stable 
pre-formed pairs. 
 We also adopt the idea in this paper. 
 However, the approximation adopted by Maly {\it et al.}, in which one of 
the Green functions in the T-matrix is renormalized and 
the other is unrenormalized, is not natural, and 
the idea of 'resonances' is artificially introduced.  
 Moreover, they have assumed the gap-like self-energy 
$ {\mit{\it \Sigma}}^{{\rm R}} (\mbox{\boldmath$k$}, \omega) \propto 
\frac{1}{\omega + \varepsilon_{\mbox{\boldmath$k$}} + {\rm i} \Gamma_{0}}$.  
 Their assumption is not self-evident for the self-consistent calculation and 
should be confirmed explicitly. Actually, we shall show later that their 
assumption cannot satisfy the self-consistency. 
 
 In this paper, on the basis of the strong coupling superconductivity, 
we make a fully self-consistent calculation to derive the character of 
the superconducting fluctuations as the 
weakly-damped pre-formed pairs 
and explicitly calculate the single particle self-energy. 

 This paper is constructed as follows. 
 In \S2, we give a model Hamiltonian and explain the basic formalism adopted 
in this paper. In \S3 and \S4, we explicitly calculate the single particle 
self-energy corresponding to the T-matrix and the self-consistent T-matrix 
approximations, respectively. 
 In \S5, we summarize the obtained results and give discussions 
on the several related issues.

\section{Basic Formalism}

 In this section we describe the basic formalism for the idea of 
the resonances of quasiparticles with the thermally excited weakly-damped 
pre-formed pairs 
on the basis of the time-dependent-Ginzburg-Landau (TDGL) expansion 
for the superconducting pairing fluctuations. 
 The pairing fluctuations have essentially different properties 
in the strong coupling case from those in the weak coupling limit.  
 We show that the strong coupling superconductivity leads to 
the anomalous normal state properties 
above the superconducting critical temperature. 
 Hereafter, we adopt the unit $\hbar=c=k_{{\rm B}}=1$. 

 We introduce the following two-dimensional model Hamiltonian which has a 
$d_{x^2-y^2}$-wave superconducting ground state, 
with High-$T_{{\rm c}}$ cuprates in mind.

\begin{eqnarray}
  \label{eq:model}
  H = \sum_{\mbox{\boldmath$k$},s} \varepsilon_{\mbox{\boldmath$k$}} 
c_{\mbox{\boldmath$k$},s}^{\dag} c_{\mbox{\boldmath$k$},s}  
+ \sum_{\mbox{\boldmath$k$},\mbox{\boldmath$k'$},\mbox{\boldmath$q$}} 
V_{\mbox{\boldmath$k$}-\mbox{\boldmath$q$}/2,
\mbox{\boldmath$k'$}-\mbox{\boldmath$q$}/2} 
c_{\mbox{\boldmath$k'$},\uparrow}^{\dag} 
c_{\mbox{\boldmath$q$}-\mbox{\boldmath$k'$},\downarrow}^{\dag} 
c_{\mbox{\boldmath$k$},\uparrow} 
c_{\mbox{\boldmath$q$}-\mbox{\boldmath$k$},\downarrow},
\end{eqnarray}

 where $ V_{\mbox{\boldmath$k$},\mbox{\boldmath$k'$}} $ is the 
$d_{x^2-y^2}$-wave separable pairing interaction, 

\begin{eqnarray}
  \label{eq:d-wave}
  V_{\mbox{\boldmath$k$},\mbox{\boldmath$k'$}} = 
g \varphi_{\mbox{\boldmath$k$}} \varphi_{\mbox{\boldmath$k'$}}, \\
  \varphi_{\mbox{\boldmath$k$}} = \cos k_{x}-\cos k_{y},
\end{eqnarray}

 where $g$ is negative. $ \varphi_{\mbox{\boldmath$k$}} $ is the 
$d_{x^2-y^2}$-wave form factor, 
and it is a constant value for the conventional $s$-wave case.

 We consider the dispersion $\varepsilon_{\mbox{\boldmath$k$}}$ given by the 
tight-binding model for a square lattice including the nearest- and 
next-nearest-neighbor hopping $t$, $t'$, respectively, 

\begin{eqnarray}
 \label{eq:dispersion}
    \varepsilon_{\mbox{\boldmath$k$}} = -2 t (\cos k_{x} +\cos k_{y}) + 
                  4 t' \cos k_{x} \cos k_{y} - \mu. 
\end{eqnarray}

 We fix the lattice constant $ a = 1$. 
 We adopt $t=0.5 \rm{eV}$ and $t'=0.45 t$. These parameters reproduce 
the Fermi surface of the typical High-$T_{{\rm c}}$ cuprates,
${\rm Y}{\rm Ba}_{2}{\rm Cu}_{3}{\rm O}_{6+\delta}$ and 
${\rm Bi}_{2}{\rm Sr}_{2}{\rm Ca}{\rm Cu}_{2}{\rm O}_{8+\delta}$. 
 We choose the chemical potential $\mu$ so that the filling $n=0.9$. 
This filling corresponds to the hole doping $\delta=0.1$.
 The Fermi surface is shown in Fig.2.

 \begin{figure}[ht]
   \begin{center}
%    \figureheight{1.0cm}   
   \epsfysize=6cm
$$\epsffile{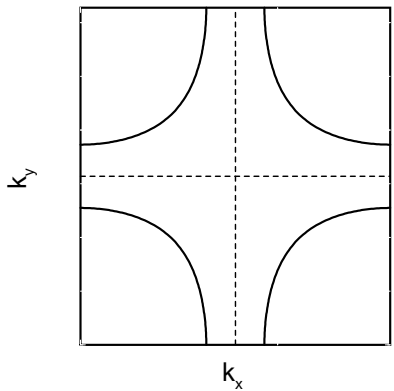}$$
%     \epsfile{file=Fig2,height=6cm}
     \caption{The Fermi surface adopted in this paper.}
     \label{fig:Fermi-surface}
   \end{center}
 \end{figure}

 Since the Brillouin zone edge acts as a natural momentum cut-off, we do not 
have to use the renormalization methods in order to remove the ultraviolet 
divergences which exist in the Fermi gas model.~\cite{rf:randeriareview}

 In reality, the origin of the pairing interaction should be considered to be 
the anti-ferromagnetic spin fluctuations. The spin fluctuations not only 
cause the pairing interaction but also affect the electronic state such as 
the momentum dependent lifetime, the momentum dependent mass enhancement 
and so on~\cite{rf:yanase}. 
 There are studies dealing with the pairing correlations obtained by 
the spin fluctuations on the basis of the fluctuation exchange (FLEX) 
approximation.~\cite{rf:dahm,rf:koikegamisuper} 
 However, we do not positively adopt these effects 
because we think that these details do not qualitatively affect the pseudogap
phenomena as a precursor of the $d_{x^{2}-y^{2}}$-wave superconductivity. 
 Indeed, from both a theory of the transport phenomena~\cite{rf:yanase} and 
the NMR experiments,~\cite{rf:NMR,rf:isida} 
it should be considered that the low frequency component of 
the anti-ferromagnetic spin fluctuations is suppressed 
in the pseudogap region. The electronic state is mainly affected by the 
low frequency component. 
 Moreover, this suppression does not contradict with our assumption that 
the pairing interaction is not suppressed in the pseudogap region, 
because the pairing interaction is mainly caused by 
the high frequency component of the spin fluctuations. 

 First of all, we consider the properties of the scattering vertex 
arising from the superconducting fluctuations, 
$\Gamma(\mbox{\boldmath$k$},\mbox{\boldmath$q$}-\mbox{\boldmath$k$}:
\mbox{\boldmath$k'$},\mbox{\boldmath$q$}-\mbox{\boldmath$k'$}:
{\rm i} \Omega_{n})$. 
 The vertex is described by the ladder diagrams (T-matrix) 
in the particle-particle interaction channel as shown in Fig.3. 

\begin{figure}[ht]
  \begin{center}
%   \figureheight{1.0cm}    
   \epsfysize=6cm
$$\epsffile{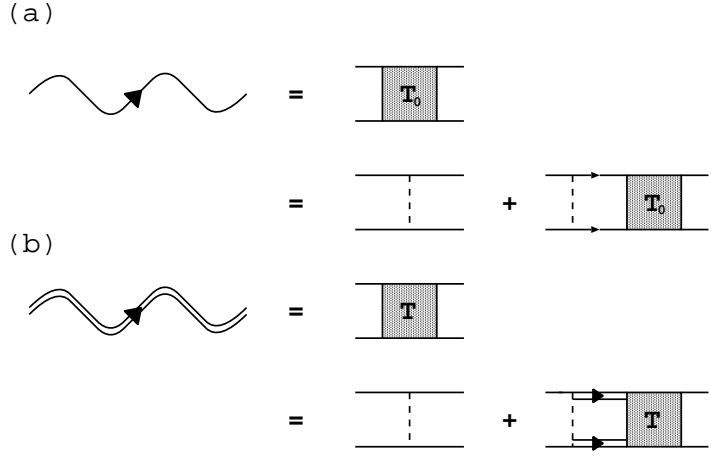}$$
%    \epsfile{file=Fig3,height=6cm}
    \caption{The scattering vertex represented by the ladder diagrams 
             in the particle-particle channel (T-matrix).  
             The dashed lines represent the attractive interaction. 
             The single and double solid lines represent the propagators of 
             the bare and renormalized fermions, respectively. 
             The single and double wavy lines represent the propagators of 
             the bare and renormalized fluctuating Cooper pairs, 
             respectively. 
             }             
    \label{fig:scattering-vertex}
  \end{center}
\end{figure}

 It is factorized into 
$\Gamma(\mbox{\boldmath$k$},\mbox{\boldmath$q$}-\mbox{\boldmath$k$}:
\mbox{\boldmath$k'$},\mbox{\boldmath$q$}-\mbox{\boldmath$k'$}:
{\rm i} \Omega_{n}) = 
\varphi_{\mbox{\boldmath$k$}-\mbox{\boldmath$q$}/2} 
t(\mbox{\boldmath$q$},{\rm i} \Omega_{n}) 
\varphi_{\mbox{\boldmath$k'$}-\mbox{\boldmath$q$}/2}$, 

 where

\begin{eqnarray}
  \label{eq:t-matrix}
  t(\mbox{\boldmath$q$},{\rm i} \Omega_{n})^{-1} & = & 
  g^{-1} + \chi_{0}(\mbox{\boldmath$q$},{\rm i} \Omega_{n}), 
\nonumber \\
  \chi_{0}(\mbox{\boldmath$q$},{\rm i} \Omega_{n}) & = & 
  T \sum_{\mbox{\boldmath$k'$},\omega_{m}} 
   {\mit{\it G}} (\mbox{\boldmath$k'$},{\rm i} \omega_{m})
   {\mit{\it G}} (\mbox{\boldmath$q$}-\mbox{\boldmath$k'$},
   {\rm i} \Omega_{n} - {\rm i} \omega_{m})
   \varphi_{\mbox{\boldmath$k'$}-\mbox{\boldmath$q$}/2}^{2}.
\end{eqnarray}

 Here, $\omega_{m} = 2 \pi (m+\frac{1}{2}) T$ and $\Omega_{n}= 2 \pi n T$ are 
the fermionic and bosonic Matsubara frequencies, respectively. 

 As a result of the analytic continuation,  
$ \chi_{0}(\mbox{\boldmath$q$}, \Omega) $ is expressed as 

\begin{eqnarray}
  \label{eq:chi0-general}
   \chi_{0}^{{\rm R}} (\mbox{\boldmath$q$}, \Omega) = 
     \sum_{\mbox{\boldmath$k$}} \int & \frac{{\rm d} \omega}{\pi} & 
     [ f(\omega-\Omega)  {\rm Im} {\mit{\it G}}^{{\rm R}} 
       (\mbox{\boldmath$q$}-\mbox{\boldmath$k$}, \Omega-\omega) 
       {\mit{\it G}}^{{\rm R}} (\mbox{\boldmath$k$}, \omega) 
\nonumber \\
   &  & - f(\omega) {\mit{\it G}}^{{\rm R}} 
       (\mbox{\boldmath$q$}-\mbox{\boldmath$k$}, \Omega-\omega) 
       {\rm Im} {\mit{\it G}}^{{\rm R}} (\mbox{\boldmath$k$}, \omega)] 
      \varphi_{\mbox{\boldmath$k$}-\mbox{\boldmath$q$}/2}^{2}.  
\end{eqnarray}

 Here, $ Z_{sc} = (1 + g \chi_{0}(\mbox{\boldmath$q$}, \Omega))^{-1} $ 
is regarded as a enhancement factor for the superconducting susceptibility
$\chi_{sc}(\mbox{\boldmath$q$},\Omega)$. 
When  $ 1 + g \chi_{0}(\mbox{\boldmath$0$},0) = 0 $, 
$\chi_{sc}(\mbox{\boldmath$0$},0)$ diverges and the superconductivity occurs. 
This is the famous Thouless criterion which is equivalent to the BCS theory 
in the weak coupling limit.~\cite{rf:Nozieres}
 
 $ t(\mbox{\boldmath$q$},\Omega) $ can be regarded 
as a propagator of the fluctuating Cooper pairs. 
The Thouless criterion corresponds to the situation in which  
$ t(\mbox{\boldmath$q$},\Omega) $ has its pole at 
$ \mbox{\boldmath$q$} = \Omega = 0 $. 

 Here, we are interested in the normal state near 
the superconducting critical point, 
where the $ Z_{sc} $ grows and the superconducting fluctuations are 
strongly enhanced. 
 Even in the weak coupling limit, this divergence affects 
the various physical quantities,~\cite{rf:AL,rf:MT} and 
the several studies based on such a framework have been given 
for High-$T_{{\rm c}}$ 
cuprates.~\cite{rf:kuboki,rf:heym,rf:randeria,rf:castro,rf:varlamov,rf:eschrig}
 However, in the strong or intermediate coupling region, 
the superconducting fluctuations more seriously affect the electronic states. 
 We will show that below. 

 Because $ t(\mbox{\boldmath$q$},\Omega) $ is strongly enhanced at 
$ \mbox{\boldmath$q$} = \Omega = 0 $ near the superconducting critical point, 
its contribution to the single particle self-energy 
$ {\mit{\it \Sigma}}(\mbox{\boldmath$k$},\omega) $ is mainly from 
the vicinity of $ \mbox{\boldmath$q$} = \Omega = 0 $.
 Therefore, we expand $ t^{-1}(\mbox{\boldmath$q$},\Omega) $ 
in the vicinity of $ \mbox{\boldmath$q$} = \Omega = 0 $.

 We describe the expansion as follows, 

\begin{eqnarray}
  \label{eq:TDGL}
   g t^{-1}(\mbox{\boldmath$q$},\Omega) = 
   t_{0} + b \mbox{\boldmath$q$}^{2} - (a_{1}+{\rm i}a_{2}) \Omega. 
\end{eqnarray}

 This expansion corresponds to the time-dependent-Ginzburg-Landau (TDGL) 
expansion up to the second order. 
 In the above description, $ t_{0} = 1 + \chi_{0}(\mbox{\boldmath$0$},0) $.  
 The other parameters $b$, $ a_{1}+{\rm i}a_{2}$ are expressed by 
the differentiation of $ \chi_{0}(\mbox{\boldmath$q$},\Omega) $ 
with respect to the momentum and frequency, respectively. 

 Later, we explicitly estimate $ \chi_{0}(\mbox{\boldmath$q$},\Omega) $ and 
the TDGL expansion parameters, 
for the non-interacting Green function (bare Green function) 
${\mit{\it G}}^{{\rm R (0)}} (\mbox{\boldmath$k$},\omega) = 
(\omega - \varepsilon_{\mbox{\boldmath$k$}} + {\rm i} \delta)^{-1} $ in \S3, 
and for the renormalized Green function  
${\mit{\it G}}^{{\rm R}} (\mbox{\boldmath$k$},\omega) = 
(\omega - \varepsilon_{\mbox{\boldmath$k$}} - 
{\mit{\it \Sigma}}^{{\rm R}} (\mbox{\boldmath$k$},\omega))^{-1} $ in \S4, 
respectively.

 These estimations correspond to the T-matrix approximation (see Fig.4(a)) 
and the self-consistent T-matrix approximation (see Fig.4(b)) 
for the single particle self-energy, respectively.
 We carry out the explicit calculation in the following sections. 

 In this section, we give an account of the general properties of the 
TDGL parameters, and phenomenologically calculate the contribution of 
the superconducting fluctuations to the single particle self-energy.

\begin{figure}[ht]
  \begin{center}
%   \figureheight{1.0cm}
   \epsfysize=8cm
$$\epsffile{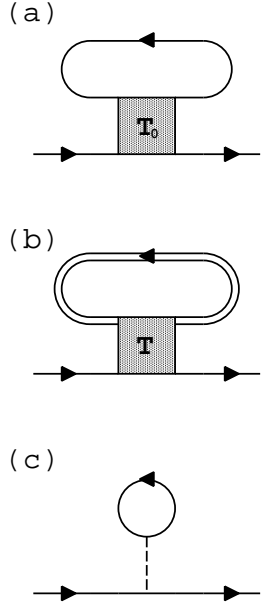}$$
%    \epsfile{file=Fig4,height=8cm}  
    \caption{The diagrams of the single particle self-energy based on 
             (a) the T-matrix approximation and 
             (b) the self-consistent T-matrix approximation, respectively.
             (c) The Hartree-Fock term which we exclude afterward. 
              We consider that this term is included 
              in the dispersion relation 
              $\varepsilon_{\mbox{\boldmath$k$}}$ from the beginning.   }
    \label{fig:selfenergy-diagram}
  \end{center}
\end{figure}

 First, we describe the expression in the weak coupling limit.
 In this case, we can use the bare Green function 
${\mit{\it G}}^{{\rm R (0)}}$.  
 $ \chi_{0}(\mbox{\boldmath$q$},\Omega) $ is 
expressed as follows: 

\begin{eqnarray}
  \label{eq:chi0}
  \chi_{0}(\mbox{\boldmath$q$},\Omega) = \sum_{\mbox{\boldmath$k$}} 
 \frac{1 - f(\varepsilon_{\mbox{\boldmath$k$}+\mbox{\boldmath$q$}/2})
         - f(\varepsilon_{-\mbox{\boldmath$k$}+\mbox{\boldmath$q$}/2})}
      {\varepsilon_{\mbox{\boldmath$k$}+\mbox{\boldmath$q$}/2}
       +\varepsilon_{-\mbox{\boldmath$k$}+\mbox{\boldmath$q$}/2}-\Omega}
 \varphi_{\mbox{\boldmath$k$}}^{2},
\end{eqnarray}

where, $f(\varepsilon)$ is the Fermi distribution function. 
Therefore, the parameters are expressed as follows:

\begin{eqnarray}
  \label{eq:parameter}
t_{0} & = & 1 + g \int {\rm d}\varepsilon 
  \frac{\tanh(\frac{\varepsilon}{2 T})}{2 \varepsilon} 
  \rho_{{\rm d}}(\varepsilon) 
  \cong |g| \rho_{{\rm d}}(0) \frac{T-T_{{\rm c}}}{T_{{\rm c}}}, 
\nonumber \\
b & = & |g| \int {\rm d}\varepsilon 
  \frac{\rho_{{\rm d}}(\varepsilon) \bar{v}_{{\rm F}}^{2}}{16 \varepsilon} 
%  \frac{1}{\varepsilon} 
  \frac{\partial^{2} f(\varepsilon)}{\partial \varepsilon^{2}}
  \cong |g| \rho_{{\rm d}}(0) \frac{7 \zeta(3)}{32 (\pi T)^{2}}  
  \bar{v}_{{\rm F}}^{2},
\nonumber \\
a_{1} & = & |g| \int {\rm d}\varepsilon 
  \frac{\tanh(\frac{\varepsilon}{2 T})}{(2 \varepsilon)^{2}}
  \rho_{{\rm d}}(\varepsilon), 
\nonumber \\
a_{2} & = & |g| \rho_{{\rm d}}(0) \frac{\pi}{8 T}, 
\end{eqnarray}

 where $ \zeta(3) $ is the Riemann's zeta function and $ \bar{v}_{{\rm F}} $ 
is the mean value of the quasiparticle's velocity on the Fermi surface.
 Here, we have defined the effective density of states 
for the $ d_{x^{2}-y^{2}} $-wave symmetry. 

\begin{eqnarray}
  \label{eq:rhod}
  \rho_{{\rm d}}(\varepsilon) = 
  \sum_{\mbox{\boldmath$k$}} \rho_{\mbox{\boldmath$k$}}(\varepsilon) 
  \varphi_{\mbox{\boldmath$k$}}^{2}, 
\end{eqnarray}

 where, $ \rho_{\mbox{\boldmath$k$}}(\varepsilon) $ is the spectral weight 
$ \rho_{\mbox{\boldmath$k$}}(\varepsilon) 
=  A_{\mbox{\boldmath$k$}}(\varepsilon) = 
- \frac{1}{\pi} {\rm Im} {\mit{\it G}}^{{\rm R}} (\mbox{\boldmath$k$},\omega) $. 

 It should be noticed that $\rho_{{\rm d}}(\varepsilon)$ is more sensitive to 
the pseudogap formation rather than the density of states 
$  \rho(\varepsilon) = 
  \sum_{\mbox{\boldmath$k$}} \rho_{\mbox{\boldmath$k$}}(\varepsilon) $.
 Because of the $ d_{x^{2}-y^{2}} $-wave-like pseudogap formation, 
the quasiparticle peak vanishes in the vicinity of $(\pi,0)$ 
where the form factor $ \varphi_{\mbox{\boldmath$k$}} $ is large.
 Although the quasiparticle peak remains in the vicinity of $(\pi/2,\pi/2)$, 
this area contributes little to $ \rho_{{\rm d}}(\varepsilon) $
because of the form factor $ \varphi_{\mbox{\boldmath$k$}} $. 
 Therefore, the pseudogap formation remarkably suppresses 
$ \rho_{{\rm d}}(\varepsilon) $.
 In this sense, the difference between the $s$-wave and the $d$-wave is small. 

 The parameter $ b $ is generally related to the coherence length $\xi$, 
$b \propto \xi^{2}$, and is very large in the weak coupling limit. 
 As the coupling becomes stronger and the critical temperature increases, 
$ b $ decreases. 
 Moreover, the pseudogap formation decreases $b$. 
 In this case, which means the short coherence length superconductor, 
the superconducting fluctuations become notable. 

 Generally, $a_{1} = 0 $ for the particle-hole symmetric case. 
However, it is non-zero in the realistic asymmetric systems. 
The general expression for $ a_{1} $ has been given 
by Ebisawa and Fukuyama.~\cite{rf:ebisawa} 
 Following them, we obtain 

\begin{eqnarray}
  \label{eq:a1}
    a_{1} \propto \frac{\partial \rho_{{\rm d}}(\varepsilon)}
{\partial \varepsilon}|_{\varepsilon=0}. 
\end{eqnarray}
 
 This quantity has been related to the Hall coefficient 
in the superconducting critical region.~\cite{rf:fukuyama}  
 It has been pointed out~\cite{rf:aronov}  
that the sign-reversal of the Hall coefficient 
under the high magnetic fields for High-$T_{{\rm c}}$ cuprates 
is not consistent with the above general expression. 

 In the weak coupling limit, $ a_{1} $ is higher order than $ a_{2} $ 
with respect to the small parameter $T_{{\rm c}}/\varepsilon_{{\rm F}}$. 
 Therefore, $ a_{1} $ is usually neglected except for the Hall coefficient. 
 However, $ a_{1} $ should not be neglected for High-$T_{{\rm c}}$ cuprates 
which is expected to be in the intermediate or strong coupling region, 
because $T_{{\rm c}}/\varepsilon_{{\rm F}}$  increases with 
the coupling constant $|g|$. 

 Furthermore, when the pseudogap opens, $ a_{2} $ remarkably decreases 
because of the suppression of the effective density of states 
$ \rho_{{\rm d}}(0) $. 
 On the other hand, the pseudogap formation does not have much effect 
on $ a_{1} $, because there is the same order contribution to $ a_{1} $ 
from the high energy part as well as from the low energy part considered 
by Ebisawa and Fukuyama (eq.~\ref{eq:a1}).  
 The expression by Ebisawa and Fukuyama is justified for the pairing 
interaction by the electron-phonon mechanism 
in which the pairing interaction is restricted to the vicinity of 
the Fermi surface. For High-$T_{{\rm c}}$ cuprates, however, 
the pairing interaction is not restricted so 
because of the non-electron-phonon mechanism. 
 Therefore, there exists the contribution from the high energy part.  
 For High-$T_{{\rm c}}$ cuprates, the contribution from the high energy part 
is mainly from the vicinity of the Van Hove singularity at $ (\pi,0) $, 
and has the same sign as that from the low energy part. 
 It is not affected by the pseudogap formation.  
 Therefore, $ a_{1} $ is not so greatly reduced by 
the pseudogap formation, and we must precisely estimate 
$ a_{1} $ using eq.~\ref{eq:chi0-general}. 

 As a result of the above discussion, we expect 
that the condition $ |a_{1}| \geq a_{2} $ is realized 
near the superconducting critical point
in the intermediate or strong coupling case. 
 This expectation is confirmed by our explicit self-consistent calculation 
in \S4. 
 This expectation means that the fluctuating Cooper pairs become propagative 
and obtain the character of the pre-formed pairs, 
although they are over-damped diffusive mode
in the conventional weak coupling theory. 

 It should be noticed that the sign of $ a_{1} $ depends on 
the band structure. 
 For High-$T_{{\rm c}}$ cuprates, $ a_{1} $ is negative. 
 This fact indicates that the pre-formed pairs are hole-like 
in High-$T_{{\rm c}}$ cuprates.

 Geshkenbein {\it et al.} phenomenologically assumed that the pre-formed pairs 
are electron-like in order to explain the sign problem of the fluctuational 
Hall coefficient.~\cite{rf:geshkenbein} 
 However, even in the strong coupling region, the sign of $a_{1}$ is negative 
and the pre-formed pairs are hole-like. 
 Therefore, we think that the assumption by Geshkenbein {\it et al.} 
is broken. 
 We consider that the simple strong coupling scenario cannot explain 
the sign problem. 
   
 In the remaining part of this section, 
we calculate the single particle self-energy 
corresponding to the one-loop diagram (Fig.5(a))
on the basis of the above TDGL expansion for the T-matrix. 
 Here, we phenomenologically define the TDGL parameters and 
use the bare Green function for simplicity. 
 Hereafter, we choose $t_{0}$ as a small parameter. 
 Therefore, our theory is an approach starting from the actual 
superconducting critical point.

\begin{figure}[ht]
  \begin{center}
%   \figureheight{1.0cm}  
   \epsfysize=6cm
$$\epsffile{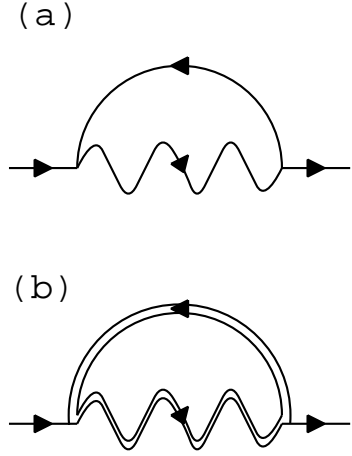}$$
%    \epsfile{file=Fig5,height=6cm}
    \caption{The single particle self-energy of the one-loop diagram.
             The wavy lines represent the propagator of the 
             fluctuating Cooper pairs described by the TDGL parameters. 
             Fig.5(a) and Fig.5(b) correspond to 
             Fig.4(a) and Fig.4(b), respectively.  }
    \label{fig:one-loop-diagram}
  \end{center}
\end{figure}

 The self-energy is given by

\begin{eqnarray}
  \label{eq:selfenergymatubara}
  {\mit{\it \Sigma}} (\mbox{\boldmath$k$}, {\rm i} \omega_{n}) = 
  T \sum_{\mbox{\boldmath$q$},{\rm i} \Omega_{m}}
  t(\mbox{\boldmath$q$},{\rm i} \Omega_{m}) 
  {\mit{\it G}} (\mbox{\boldmath$q$}-\mbox{\boldmath$k$}, 
   {\rm i} \Omega_{m} - {\rm i} \omega_{n})
  \varphi_{\mbox{\boldmath$k$}-\mbox{\boldmath$q$}/2}^{2} .    
\end{eqnarray}

 After the analytic continuation, we obtain

\begin{eqnarray}
  \label{eq:selfenergyreal}
  {\mit{\it \Sigma}}^{{\rm R}} (\mbox{\boldmath$k$}, \omega) & = & 
  \sum_{\mbox{\boldmath$q$}} \int \frac{{\rm d} \Omega}{\pi} 
  [ b(\Omega) {\rm Im} t(\mbox{\boldmath$q$},\Omega) 
   {\mit{\it G}}^{{\rm A}} (\mbox{\boldmath$q$}-\mbox{\boldmath$k$}, 
    \Omega-\omega)
   -f(\Omega) t(\mbox{\boldmath$q$},\Omega+\omega) 
   {\rm Im} {\mit{\it G}}^{{\rm R}} (\mbox{\boldmath$q$}-\mbox{\boldmath$k$}, 
    \Omega) ] 
   \varphi_{\mbox{\boldmath$k$}-\mbox{\boldmath$q$}/2}^{2}, 
\nonumber  \\ 
   {\rm Im} {\mit{\it \Sigma}}^{{\rm R}} (\mbox{\boldmath$k$}, \omega) & = & 
   -\sum_{\mbox{\boldmath$q$}} \int \frac{{\rm d} \Omega}{\pi} 
   [ b(\Omega+\omega)+f(\Omega) ] {\rm Im} 
     t(\mbox{\boldmath$q$},\Omega+\omega) 
   {\rm Im} {\mit{\it G}}^{{\rm R}} (\mbox{\boldmath$q$}-\mbox{\boldmath$k$}, 
    \Omega)
   \varphi_{\mbox{\boldmath$k$}-\mbox{\boldmath$q$}/2}^{2} ,
\end{eqnarray}

 where $b(\Omega)$ is the Bose distribution function.
 In the conventional Fermi liquid theory, the factor 
$ [ b(\Omega+\omega)+f(\Omega) ] $ gives the common relation 
${\rm Im}{\mit{\it \Sigma}}^{{\rm R}}(\mbox{\boldmath$k$},\omega) 
\propto \omega^{2} +(\pi T)^{2}$. 
 However, when the strong fluctuations exist, 
the strong $ \mbox{\boldmath$q$} $- and $ \Omega $-dependence of 
$ t(\mbox{\boldmath$q$},\Omega) $ give rise to the anomalous features 
as we show below. 

 Assuming $ |a_{1}| \gg a_{2} $, we describe the imaginary part of 
the T-matrix as 

\begin{eqnarray}
  \label{eq:Imt(q,w)}
  {\rm Im} t(\mbox{\boldmath$q$},\Omega) = 
  g \frac{\pi}{a_{1}} \delta(\Omega - \Omega_{\mbox{\boldmath$q$}})
\end{eqnarray}

 Here, we have defined $ \Omega_{\mbox{\boldmath$q$}} = 
(t_{0} + b \mbox{\boldmath$q$}^{2})/a_{1} $. 
 This quantity corresponds to the dispersion relation of the pre-formed pairs.
 Because $ a_{1} $ is negative in our case, 
$ \Omega_{\mbox{\boldmath$q$}} < 0 $. 
 Therefore, the pre-formed pairs have not electron-like 
but hole-like character. 

 Substituting eq.~\ref{eq:Imt(q,w)} into eq.~\ref{eq:selfenergyreal}, 
we obtain 
 
\begin{eqnarray}
  \label{eq:selfenergydelta}
  {\rm Re} {\mit{\it \Sigma}}^{{\rm R}} (\mbox{\boldmath$k$}, \omega) & = &  
  \sum_{\mbox{\boldmath$q$}} 
  g [ \frac{1}{a_{1}} b(\Omega_{\mbox{\boldmath$q$}}) 
      {\rm Re} {\mit{\it G}}^{{\rm R}} 
                (\mbox{\boldmath$q$}-\mbox{\boldmath$k$}, 
                 \Omega_{\mbox{\boldmath$q$}}-\omega)
\nonumber \\
  & &  -  \int \frac{{\rm d} \Omega}{\pi} f(\Omega-\omega) 
       \frac{1}{t_{0}+b \mbox{\boldmath$q$}^{2}-a_{1} \Omega} 
       {\rm Im} {\mit{\it G}}^{{\rm R}} 
       (\mbox{\boldmath$q$}-\mbox{\boldmath$k$},\Omega-\omega)]
  \varphi_{\mbox{\boldmath$k$}-\mbox{\boldmath$q$}/2}^{2} , 
\nonumber \\
  {\rm Im} {\mit{\it \Sigma}}^{{\rm R}} (\mbox{\boldmath$k$}, \omega) & = &  
  - \sum_{\mbox{\boldmath$q$}} 
    \frac{g}{a_{1}} [b(\Omega_{\mbox{\boldmath$q$}}) + 
                     f(\Omega_{\mbox{\boldmath$q$}}-\omega)]
    {\rm Im} {\mit{\it G}}^{{\rm R}} (\mbox{\boldmath$q$}-\mbox{\boldmath$k$}, 
      \Omega_{\mbox{\boldmath$q$}}-\omega) 
    \varphi_{\mbox{\boldmath$k$}-\mbox{\boldmath$q$}/2}^{2} .
\end{eqnarray}

 We can see from the above expressions that the singular contribution 
in the small $t_{0}$ limit arises from the terms proportional to 
$b(\Omega_{\mbox{\boldmath$q$}})$. 
 Therefore, we estimate the singular terms 
for the linearized bare Green function 
$ {\mit{\it G}}^{{\rm (0)R}} (\mbox{\boldmath$q$}-\mbox{\boldmath$k$},\omega) 
= (\omega - \varepsilon_{\mbox{\boldmath$k$}} + 
v_{\mbox{\boldmath$k$}} \mbox{\boldmath$q$} + {\rm i} \delta)^{-1} $. 
 Because the singular contribution arises from the vicinity of 
$\mbox{\boldmath$q$}=0$, we restrict the \mbox{\boldmath$q$}-sum to the region 
$|\Omega_{\mbox{\boldmath$q$}}| \leq T $, and use the approximate relation 
$b(\Omega_{\mbox{\boldmath$q$}}) \sim 
\frac{T}{\Omega_{\mbox{\boldmath$q$}}} $. 
 Since only the small region in the vicinity of $\mbox{\boldmath$q$}=0$ 
contributes to the self-energy, we can neglect the 
$\mbox{\boldmath$q$}$-dependence of the form factor 
$ \varphi_{\mbox{\boldmath$k$}-\mbox{\boldmath$q$}/2} $. 
 We exactly calculate the \mbox{\boldmath$q$}-sum. 
 However, the results are very complicated. 
 Therefore, we show the approximate results as follows

\begin{eqnarray}
  \label{eq:selfenergy}
  {\rm Re} {\mit{\it \Sigma}}^{{\rm R}} (\mbox{\boldmath$k$}, \omega) & = & 
  \left\{ 
   \begin{array}{ll} 
    - g \varphi_{\mbox{\boldmath$k$}}^{2} \frac{T}{4 \pi b} \frac{1}{\alpha} 
      \log [\frac{a_{1}}{t_{0}} {\rm min}
         (T,\alpha,\alpha^{2} b/|a_{1}| v_{\mbox{\boldmath$k$}}^{2})] 
  & (|\alpha| \gg \frac{t_{0}}{|a_{1}|}) \\
    - g \varphi_{\mbox{\boldmath$k$}}^{2} 
      \frac{T}{2 \pi a_{1} v_{\mbox{\boldmath$k$}}^{2}} 
      (\frac{a_{1}}{t_{0}} \alpha - 1) 
  & (|\alpha| \sim \frac{t_{0}}{|a_{1}|}) ,
   \end{array}
  \right. 
\\
   {\rm Im} {\mit{\it \Sigma}}^{{\rm R}} (\mbox{\boldmath$k$}, \omega) & = &  
  \left\{ 
   \begin{array}{ll} 
    g \varphi_{\mbox{\boldmath$k$}}^{2} \frac{T}{4 b} 
      (\alpha^{2} + \frac{v_{\mbox{\boldmath$k$}}^{2}}{b} t_{0})^{-\frac{1}{2}} 
  & (|\alpha| \leq T) \\
    0
  & (\alpha \geq \frac{|a_{1}| v_{\mbox{\boldmath$k$}}^{2}}{4 b} 
     - \frac{t_{0}}{|a_{1}|}) ,
   \end{array}
  \right. 
\end{eqnarray}

 where we have defined $ \alpha = \omega + \varepsilon_{\mbox{\boldmath$k$}} $.

 We show the typical features of 
$ {\rm Re} {\mit{\it \Sigma}}^{{\rm R}} (\mbox{\boldmath$k$}, \omega) $,
$ {\rm Im} {\mit{\it \Sigma}}^{{\rm R}} (\mbox{\boldmath$k$}, \omega) $ and 
$ \pi A(\mbox{\boldmath$k$}, \omega) = -  
{\rm Im} {\mit{\it G}}^{{\rm R}} (\mbox{\boldmath$k$}, \omega) $ in Fig.6. 

\begin{figure}[p]
  \begin{center}
%   \figureheight{1.0cm}  
   \epsfysize=6cm
$$\epsffile{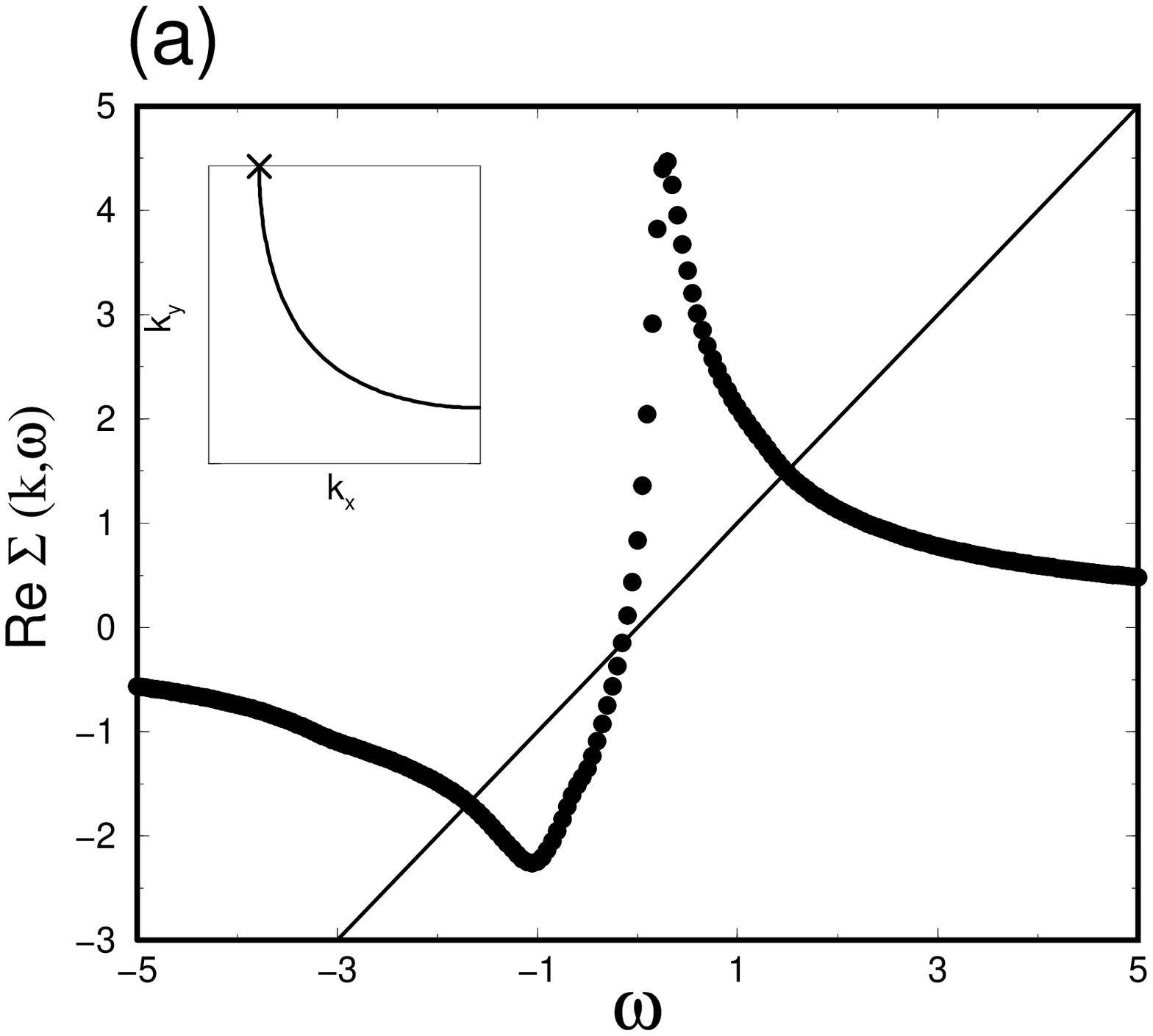}$$
   \epsfysize=6cm
$$\epsffile{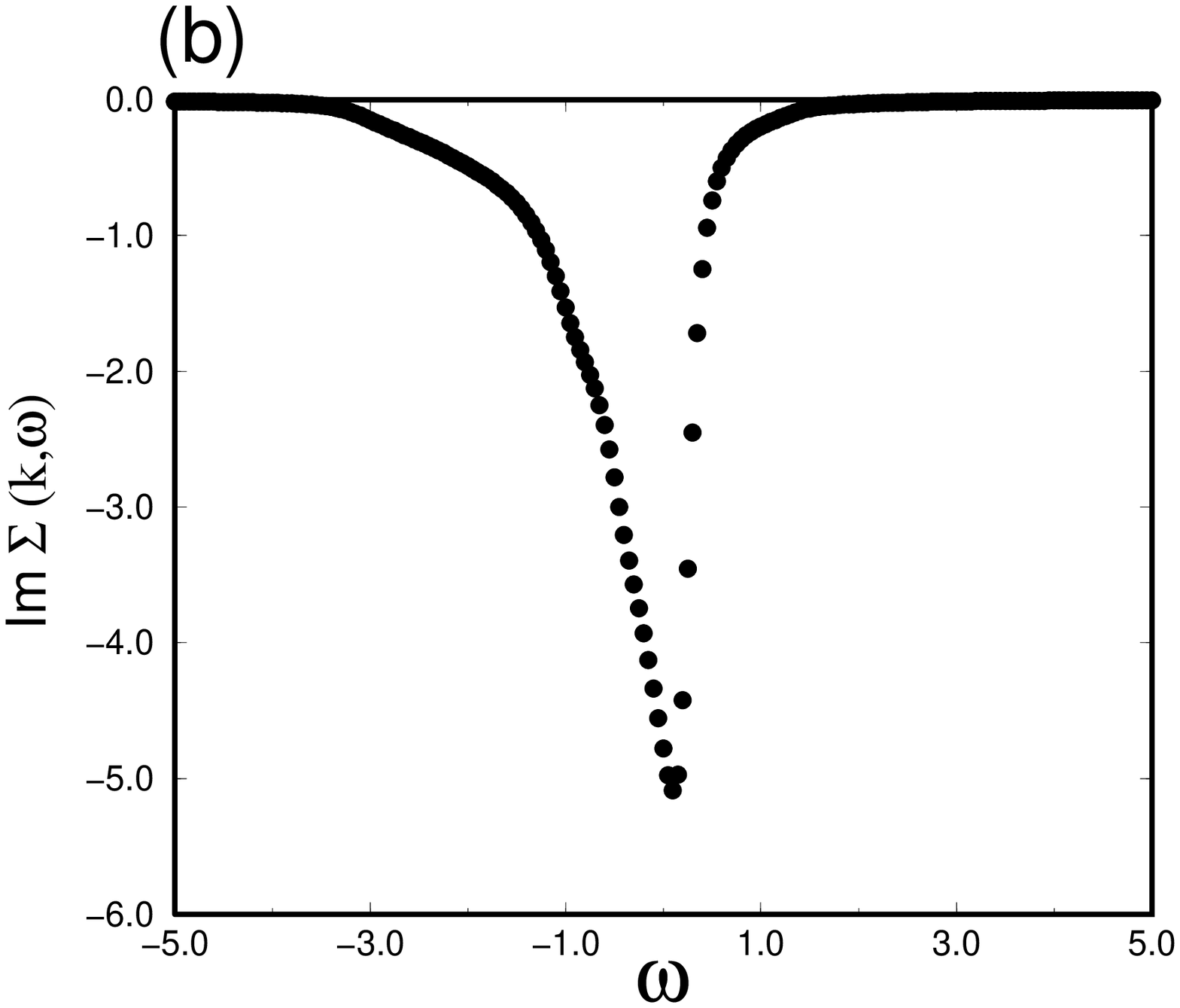}$$
   \epsfysize=6cm
$$\epsffile{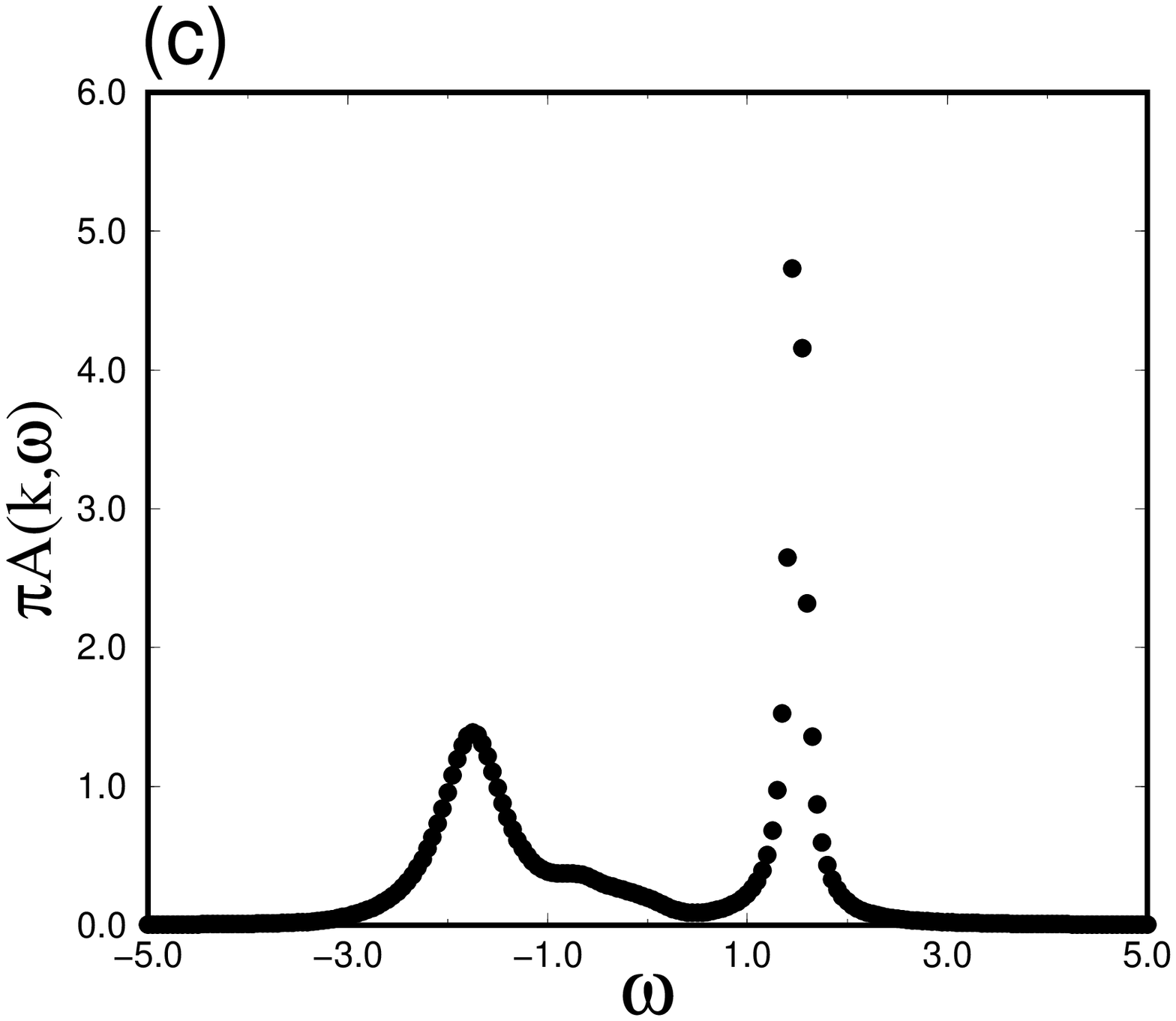}$$
%    \epsfile{file=Fig6a,height=6cm}
%    \epsfile{file=Fig6b,height=6cm}
%    \epsfile{file=Fig6c,height=6cm}
    \caption{(a) The real part of the self-energy given by 
             the one loop diagram in Fig.5(a). 
             The solid line shows the condition 
             $ \omega - \varepsilon_{\mbox{\boldmath$k$}} - 
             {\rm Re} {\mit{\it \Sigma}}^{{\rm R}} 
             (\mbox{\boldmath$k$}, \omega) 
             = 0 $. 
             (b) The imaginary part of the self-energy. 
             (c) The spectral weight. 
             For all figures, the $\mbox{\boldmath$k$}$-point 
             is chosen on the Fermi surface near $(0,\pi)$. 
             (see inset in (a)). 
             Here, the coupling constant and the TDGL parameters are 
             phenomenologically chosen 
             as $g=-2.0$, $a_{1}=-1.0$, $a_{2}=0.2$, 
             $b=0.2$, $t_{0}=0.03$, $T=0.2$.   }
    \label{fig:one-loop-selfenergy}
  \end{center}
\end{figure}

 It is notable that the real part of the self-energy has the positive slope 
in the vicinity of $ \alpha = 0 $, 
and the imaginary part of the self-energy has the sharp peak 
at $ \alpha = 0 $ in its absolute value. 
 The both features are anomalous 
compared with the conventional Fermi liquid theory. 
 These anomalous features of the single particle self-energy should be 
regarded as the effects of the resonances of quasiparticles with 
the weakly-damped thermally excited pre-formed pairs. 
 Such drastic phenomena take place on the condition 
$|\Omega_{\mbox{\boldmath$0$}}| \ll T $. 
 Of course, the characteristics of the strong coupling superconductivity are 
reflected on the properties of the TDGL parameters.  
 As the coupling constant $|g|$ increases and 
the critical temperature increases, the effects of the resonances 
become remarkable. 

 Although we have used the assumption $ |a_{1}| \gg a_{2} $ 
in the above analytic calculation, 
the results change little even in the more gentle condition 
$ |a_{1}| \sim a_{2} $. 
 Furthermore, these features do not change qualitatively 
even in the strongly diffusive region $ |a_{1}| \ll a_{2} $. 
 However, the absolute value of the self-energy is considerably small 
in the diffusive case. 
 Therefore, these effects are very small in the weak coupling case.

 It should be noticed that the self-energy essentially has 
the asymmetric structure, which 
naturally arises from the hole-like features of the pre-formed pairs. 
 In particular, the imaginary part has the long tail 
on the negative frequency side. 

 Norman {\it et al.}~\cite{rf:norman2} have used the assumption 
$ {\mit{\it \Sigma}}^{{\rm R}} (\mbox{\boldmath$k$}, \omega) \propto 
\frac{1}{\omega + \varepsilon_{\mbox{\boldmath$k$}} + {\rm i} \Gamma_{0}}$ 
in order to explain the data of the ARPES experiments. 
 Our results are qualitatively consistent with their assumption. 
 However, we shall point out in \S4 that the slight break down of 
the assumption which originates from the asymmetric structure 
is important for the self-consistent solution. 
% However, also in the self-consistent calculation, the assumption is 
%reproduced in a rough estimate. 

 The quantity $ A(\mbox{\boldmath$k$}, \omega) $ corresponds to 
the spectral weight. 
 According to Fig.6(a), there are three solutions of the equation 
$ \omega - \varepsilon_{\mbox{\boldmath$k$}} - 
{\rm Re} {\mit{\it \Sigma}}^{{\rm R}} (\mbox{\boldmath$k$}, \omega) = 0 $.
 However, since the imaginary part of the self-energy is extremely large 
in the vicinity of $\alpha = 0$, the middle solution in Fig.6(a) disappears. 
 Therefore, the spectral weight has the gap-like two peak structure at 

\begin{eqnarray}
  \label{eq:gap}
  E & = & \pm \sqrt{\varepsilon_{\mbox{\boldmath$k$}}^{2} 
                    + \Delta_{\mbox{\boldmath$k$}}^{2}} ,
\\
  \Delta_{\mbox{\boldmath$k$}}^{2} & = & 
  |g| \varphi_{\mbox{\boldmath$k$}}^{2} \frac{T}{4 \pi b} 
               \log [T/\frac{t_{0}}{a_{1}}] .
\end{eqnarray}

 This spectrum is similar to that of the conventional superconducting state.
 Although the gap amplitude is related to the condensed 
Cooper pairs in the superconducting state, here, 
$\Delta_{\mbox{\boldmath$k$}}^{2}$ is proportional to the number of 
the thermally excited pre-formed pairs 
$n_{{\rm B}} \propto \sum_{\mbox{\boldmath$q$}} 
b(\Omega_{\mbox{\boldmath$q$}})$. 
 
 Furthermore, the spectral weight also has the asymmetric structure.  
 The spectrum has the relatively broad peak on the negative frequency side and 
the sharp peak on the positive frequency side.
 Randeria {\it et al.}~\cite{rf:randeria2} have proposed the assumption of 
the particle-hole symmetry for the spectral weight, 
and actually Norman {\it et al.}~\cite{rf:norman} 
carried out the analysis for the ARPES data 
on the basis of the assumption. 
 We consider that this analysis is not so wrong qualitatively.  
However, our calculation shows that this assumption is essentially broken 
reflecting the hole-like character of the pre-formed pairs. 
 Indeed, the asymmetry plays an important role in 
the self-consistent calculation in \S4. 

 We shall numerically carry out the explicit calculation 
for the TDGL parameters 
and the single-particle self-energy corresponding to the T-matrix (in \S3) and 
the self-consistent T-matrix (in \S4) approximations 
in the following sections. 

 It is notable that the effects of the resonances described above are 
extraordinary small and can not be seen in the weak coupling limit, 
where $b$ is large and $ |a_{1}| \ll a_{2} $ 
because of the small $T_{{\rm c}}$. 
 Therefore, the effects of the resonances are the characteristics of 
the strong coupling superconductivity. 

 However, as we described above, the anomalous features, such as 
the relatively large imaginary part of the self-energy 
in the vicinity of $\alpha=0$ and so on, 
exist even in the weak coupling case. Consequently, the density of states 
at the Fermi level are slightly suppressed even in the weak coupling case. 
 This effect has already been discussed by the other authors 
on the basis of the conventional weak coupling formalism.
~\cite{rf:heym,rf:randeria,rf:castro,rf:varlamov,rf:eschrig}
 In particular, Eschrig {\it et al.}~\cite{rf:eschrig} have shown that 
the NMR experiments for the optimally-doped cuprates are understood 
on the basis of the conventional $d$-wave superconducting fluctuation theory. 
 Also in their theory, the suppression of the density of states plays an 
essential role. 
 This effect appears also in our theory. However, it is more remarkable and 
leads to the gap-like structure of the spectral weight 
in the strong coupling case. In this sense, our theory is 
a natural extension of these conventional weak coupling theories. 

 It should be noticed that 
there is a logarithmic singularity of the self-energy and 
$\Delta_{\mbox{\boldmath$k$}}$ near the critical point, 
reflecting the singularity of two-dimensional systems. 
 Strictly speaking, it leads to the fact $T_{{\rm c}} = 0$. 
 This is a natural result which satisfies the Marmin-Wagner's theorem. 
 The Nozi$\grave{{\rm e}}$res and Schmitt-Rink formalism 
for two-dimensional systems~\cite{rf:schmitt-rink,rf:tokumitu} has 
the similar singularity, and the critical temperature 
$T_{{\rm c}} = 0$. 

 However, we consider that there is not such a singularity 
in the realistic layered systems. 
 Although the quasi-two-dimensionality enhances the effects of 
the fluctuations in the layered systems, 
the weak three-dimensionality is sure to remove 
the singularity.~\cite{rf:randeria} 
 
 Here, we comment on the Nozi$\grave{{\rm e}}$res and Schmitt-Rink 
formalism. 
 The Nozi$\grave{{\rm e}}$res and Schmitt-Rink theory takes account of 
the shift of the chemical potential by the creation of the bosonic particles, 
and decides the critical temperature $T_{{\rm c}}$ 
using the Thouless criterion for the shifted chemical potential. 
 It should be noticed that for High-$T_{{\rm c}}$ cuprates 
the chemical potential shifts upward because the pre-formed pairs are 
not electron-like, but hole-like.  
It is the opposite direction compared to the Fermi gas model or 
the low density lattice model. (see Fig. 7.) 
 The upward shift seems to be natural because the density of states decreases 
in that direction. 

\begin{figure}[ht]
  \begin{center}
%   \figureheight{1.0cm}  
   \epsfysize=10cm
$$\epsffile{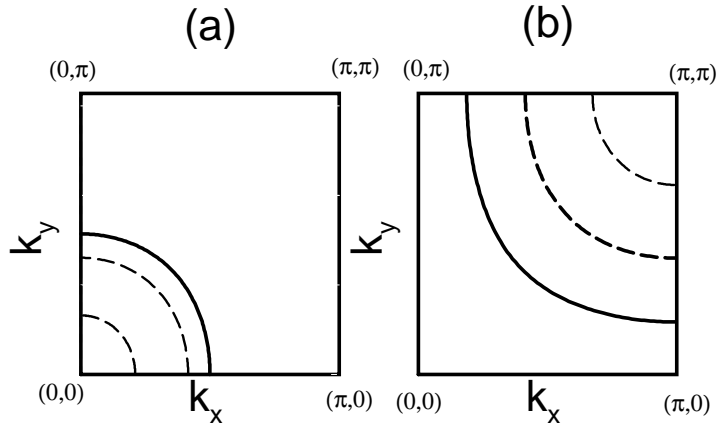}$$
%    \epsfile{file=Fig7,height=10cm}
    \caption{The schematic figures of the transformation of the Fermi surface 
             by the chemical potential shift based on the 
             Nozi$\grave{{\rm e}}$res and Schmitt-Rink theory.
             The solid line shows the original Fermi surface, and 
             the long-dashed line shows the shifted Fermi surfaces. 
             (a) The case of the Fermi gas model, or 
                 the low density lattice model. 
                 The chemical potential shifts downward.
             (b) The case of High-$T_{{\rm c}}$ cuprates. 
                 The chemical potential shifts upward.  
                 This upward shift reflects the hole-like properties of the 
                 pre-formed pairs in High-$T_{{\rm c}}$ cuprates. }
    \label{fig:chemical-potential-shift}
  \end{center}
\end{figure}

 This formalism is justified in the low density limit. 
 However, High-$T_{{\rm c}}$ cuprates should be regarded 
as rather high density limit, 
because they are the lattice systems near the half-filling. 
 Therefore, we need the extended calculation as is carried out in this paper.  

 In the low density limit, a strong attractive interaction easily creates 
the bosonic pre-formed pairs which can move almost freely, 
and the chemical potential shifts remarkably. 
 There may be no serious effect on the normal state fermion system 
except for the chemical potential shift. 
 This insight is suggested by the fact that the self-energy calculated 
in this section is proportional to 
$ n_{{\rm B}} $ in rough estimate. 
 $ n_{{\rm B}} $ can not become so large in the low density systems. 
 Therefore, the effects of the resonances described above cannot be seen 
in the low density limit. 
 On the other hand, in high density systems, 
the effects of the resonances occur in the 
fermion systems before the chemical potential shifts remarkably. 
 Therefore, although the chemical potential actually shifts also 
in the high density systems, the shift is not a dominant effect. 
 The chemical potential shift is included in our calculation. However,  
we will not pay attention to it in the following sections.

\section{Lowest Order Calculation}

 In this section we explicitly calculate the single particle self-energy 
on the basis of the formalism described in the previous section. 

 First, we calculate the T-matrix around $ \mbox{\boldmath$q$} = \Omega = 0 $, 
for the non-interacting Green function  
$ {\mit{\it G}}^{{\rm R (0)}} (\mbox{\boldmath$k$},\omega) = 
(\omega - \varepsilon_{\mbox{\boldmath$k$}} + {\rm i} \delta)^{-1} $, 
using eq.~\ref{eq:chi0-general}. 
 We expand the reciprocal of the T-matrix, 
$ t(\mbox{\boldmath$q$}, \Omega)^{-1} $ 
and estimate the TDGL expansion parameters. 
 Of course, the results are equivalent to eq.~\ref{eq:parameter}.

 Secondly, we calculate the single particle self-energy corresponding to the 
diagram shown in Fig.5(a), using eq.~\ref{eq:selfenergyreal}. 
 This calculation corresponds to the T-matrix approximation (Fig.4(a)).
 We exclude the trivial Hartree-Fock term corresponding to the diagram shown 
in Fig.4(c). We consider that this term is included in the dispersion relation 
$\varepsilon_{\mbox{\boldmath$k$}}$ from the beginning. 
 This exclusion makes no difference in the strongly fluctuating region. 
 
 Because we use the non-interacting Green function, 
the superconducting critical temperature $T_{{\rm c}}$ is the same as that 
obtained by the BCS mean field theory, $T_{{\rm MF}}$. 
 Therefore, our calculation in this section is 
carried out above $T_{{\rm MF}}$. 

 We show the results in Fig.8.

 \begin{figure}[ht]
%  \figureheight{1.0cm}
   \epsfysize=6cm
$$\epsffile{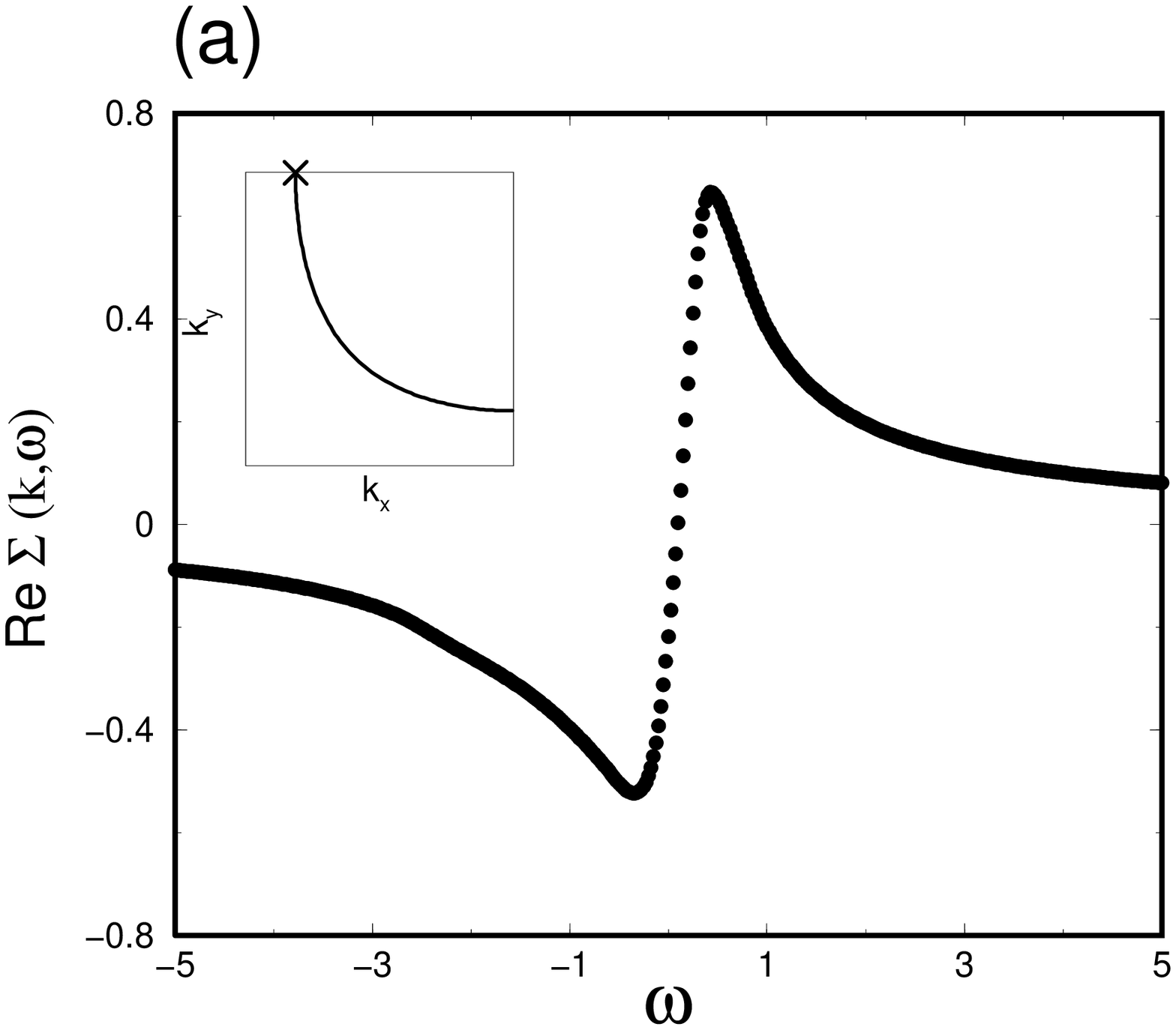}$$
   \epsfysize=6cm
$$\epsffile{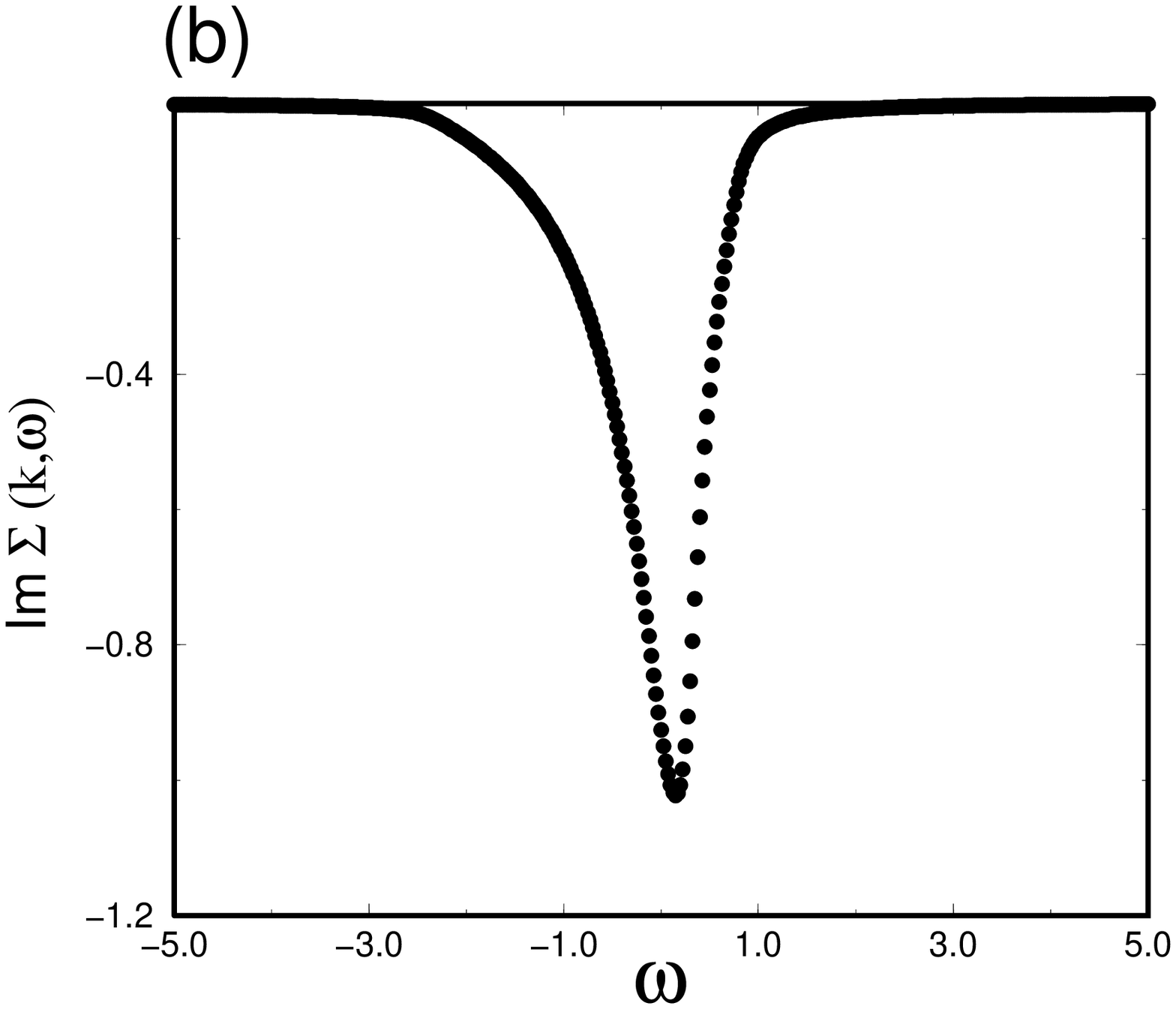}$$
%   \epsfile{file=Fig8a,height=6cm}
%   \epsfile{file=Fig8b,height=6cm}
   \begin{center}
    \caption{The single particle self-energy on the Fermi surface 
              near $(0,\pi)$ obtained by the lowest order calculation. 
              (a)The real part. (b)The imaginary part.
              Here, $\mbox{\boldmath$k$}=(0.589,\pi)$, $g=-1.0$, and $T=0.21$.
              The $\mbox{\boldmath$k$}$-point is shown in the inset. 
              Here, $T_{{\rm c}}=T_{{\rm MF}}=0.185$. }
     \label{fig:T-matrix-selfenergy}
   \end{center}
 \end{figure}

 The self-energy shows the same features as we have calculated analytically 
in \S2.
 Although $ |a_{1}| \leq a_{2} $ is always realized in this calculation, 
the effects of the resonances clearly appear. 

 The spectral weight is shown in Fig.9 for various temperatures. 

 \begin{figure}[ht]
%  \figureheight{1.0cm}
   \begin{center}
   \epsfysize=6cm
$$\epsffile{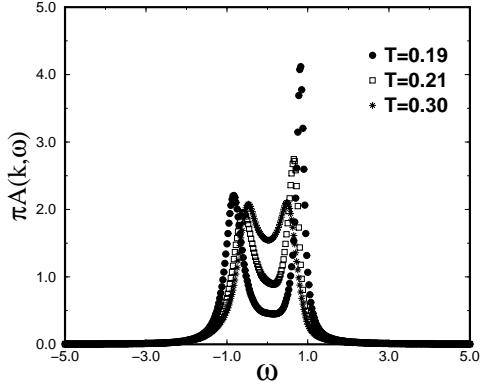}$$
%    \epsfile{file=Fig9,height=6cm}
    \caption{The spectral weight for various temperatures 
              $T=0.19$(circles), $T=0.21$(squares), $T=0.30$(stars).
              The other parameters are the same as those in Fig.8.   }
     \label{fig:T-matrix-temperature}
   \end{center}
 \end{figure}

 The spectral weight has the clear double peak gap-like structure 
in the low temperature region. As the temperature increases, the gap structure 
is filled up. 
 Thus, the quasiparticle feature of the normal Fermi liquid theory realized 
at the higher temperature becomes unstable near the superconducting 
transition temperature $T_{{\rm c}}=T_{{\rm MF}}=0.185$, where 
the superconducting fluctuations are strong. As a result, 
the resonance effects produce the gap-like double peak structure.
  
 The momentum dependence of the spectral weight is shown in Fig.10.

 \begin{figure}[ht]
%  \figureheight{1.0cm}
   \begin{center}
   \epsfysize=6cm
$$\epsffile{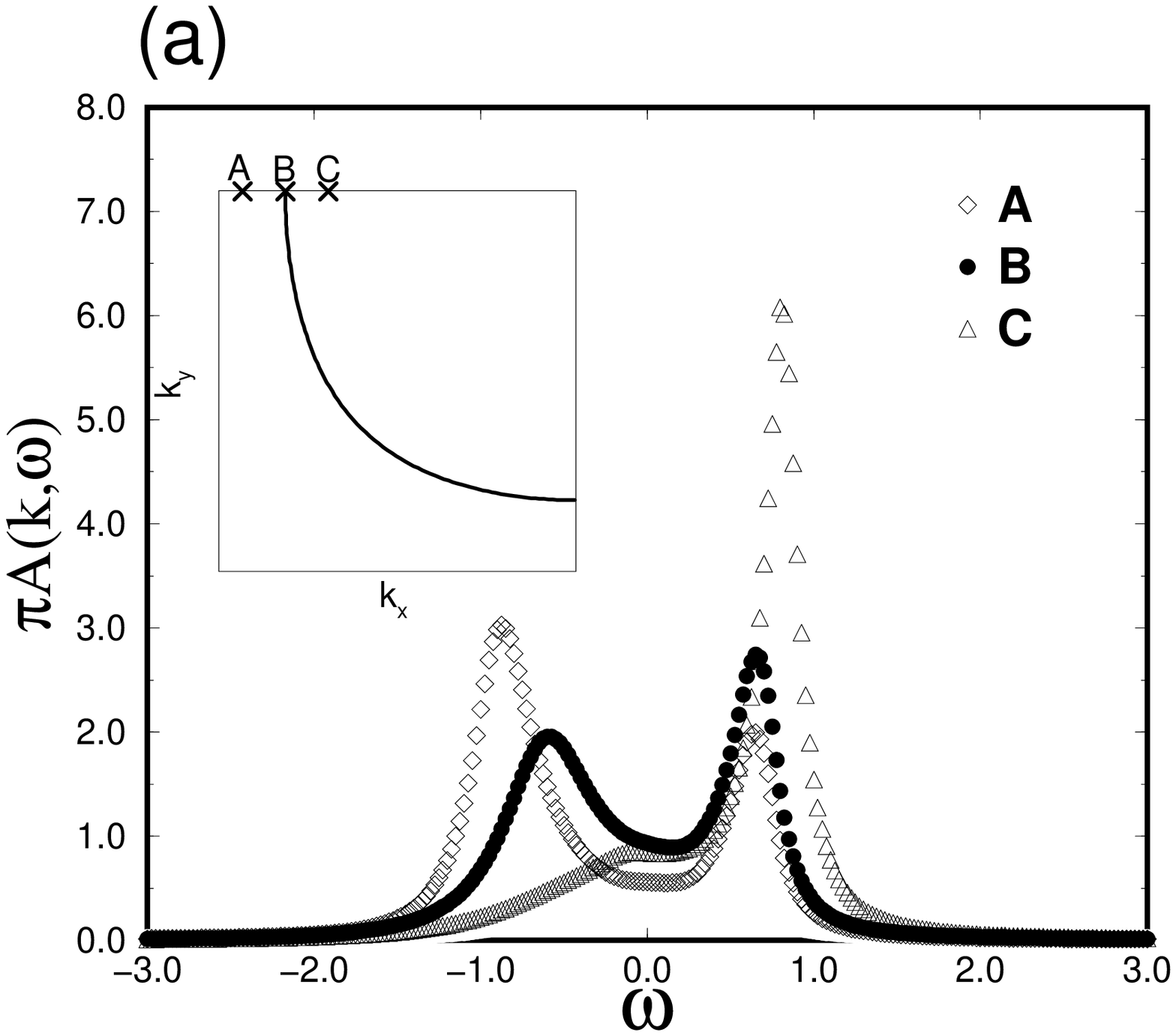}$$
   \epsfysize=6cm
$$\epsffile{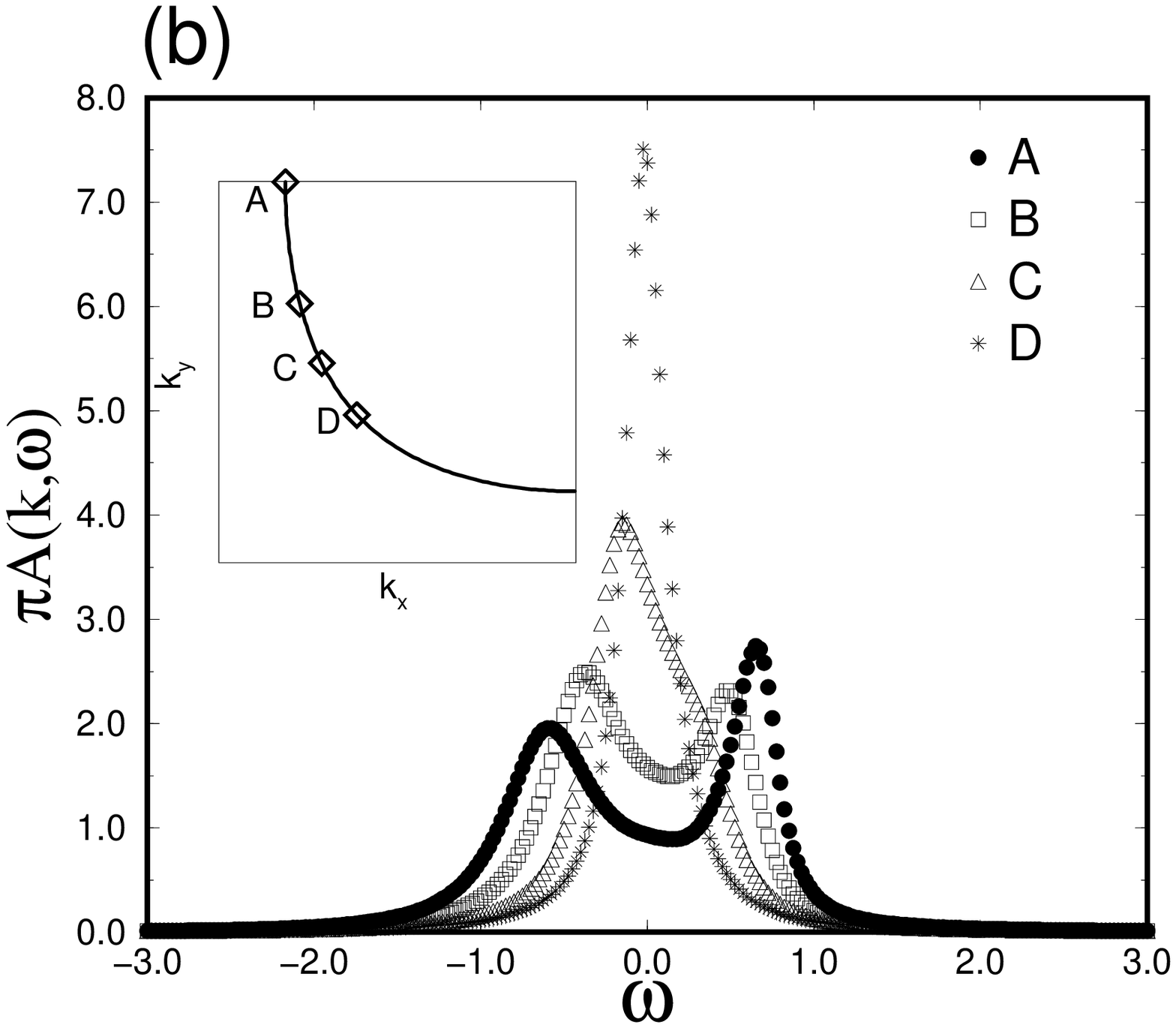}$$
%    \epsfile{file=Fig10a,height=6cm}
%    \epsfile{file=Fig10b,height=6cm}
    \caption{The momentum dependence of the spectral weight 
             (a)across the Fermi surface near $(0,\pi)$ 
             (b)along the Fermi surface. 
             The respective $\mbox{\boldmath$k$}$-points
             are shown in the inset. 
             The other parameters are the same as those in Fig.8. }
     \label{fig:T-matrix-k}
   \end{center}
 \end{figure}

 We can see that the amplitude of the gap is large in the vicinity of 
$(0,\pi)$, and decreases as the momentum approaches $(\pi/2,\pi/2)$. 
 Furthermore, the single peak structure of quasiparticles exists 
in the vicinity of $(\pi/2,\pi/2)$. 
 These features are consistent with the ARPES experiments.~\cite{rf:norman} 
 It is mainly because of the $ d_{x^{2}-y^{2}} $-wave form factor 
$ \varphi_{\mbox{\boldmath$k$}} $.  
 Thus, it is naturally led from our formalism that the pseudogap has 
the same shape as that of the superconducting gap. 
 Since the attractive interaction is small in the vicinity of $(\pi/2,\pi/2)$, 
quasiparticles are not strongly affected there 
by the superconducting fluctuations.

 The density of states $ \rho(\varepsilon) $ is shown in Fig.11.

 \begin{figure}[ht]
   \begin{center}
%    \figureheight{1.0cm} 
   \epsfysize=6cm
$$\epsffile{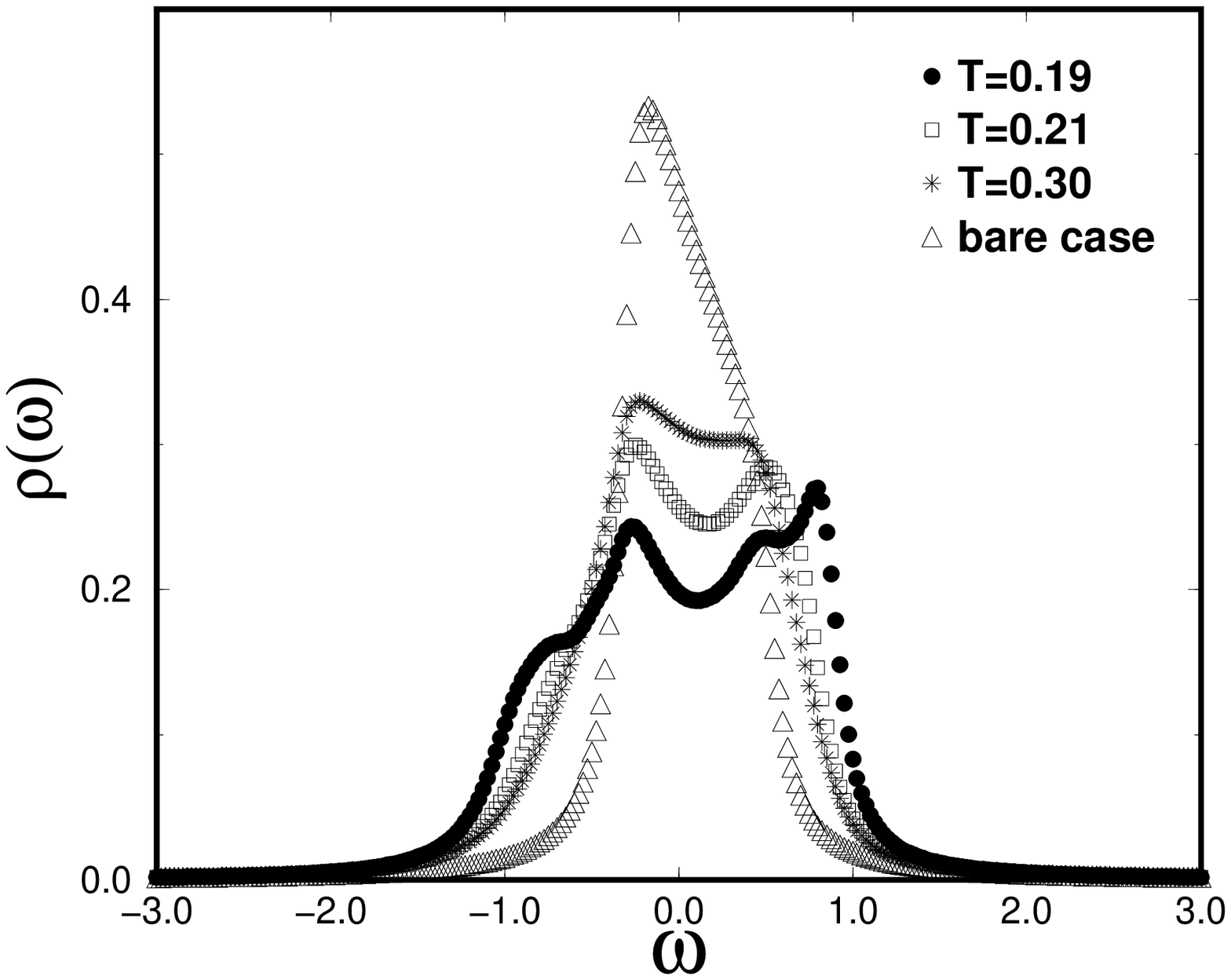}$$
%    \epsfile{file=Fig11,height=6cm}
     \caption{The density of states $ \rho(\varepsilon) $ 
              for various temperatures. 
              The parameters and the marks are the same as those in Fig.8. 
              The triangles show the non-interacting density of states.
              Since we calculate only in the vicinity of the Fermi surface, 
              the peak position is not exact. 
              The suppression of the density of states at the low energy 
              is essential.      
              }
     \label{fig:T-matrix-dos}
   \end{center}
 \end{figure}

 We calculate the self-energy only for the $\mbox{\boldmath$k$}$-points in the 
vicinity of the Fermi surface. 
 Because the quasiparticles far from the Fermi surface are not so much 
affected by the superconducting fluctuations, we do not pay attention to 
them.   
 We show the density of states in Fig.11, obtained by summing up the momentum 
on the calculated area.  
 Therefore, the peak position is artificial and depends on the summed area. 
 Only the suppression around the Fermi energy is essential.

 There exists a gap structure also in the density of states.
 However, because of the peak due to the quasiparticles 
around $(\pi/2,\pi/2)$, 
the gap structure of the density of states is not so clear as that in the 
superconducting states, or that of the spectral weight in the vicinity of 
$(0,\pi)$.  
 These results are consistent with the tunneling spectroscopy 
experiments.~\cite{rf:renner}

 Thus, most of the features related to the pseudogap phenomena are understood 
from the above lowest order calculation. 
 From the above results, we conclude at least that the strong coupling 
superconductivity makes the Fermi liquid state unstable near the mean field 
superconducting critical temperature $T_{{\rm MF}}$.   

  As we have pointed out before, the above calculation is applicable to the 
higher temperature than $T_{{\rm MF}}$. 
 In practice, the effects of the superconducting fluctuations remarkably 
suppress the critical temperature. 
 Indeed, we have an interest in the temperature region in which 
the superconductivity is suppressed by the fluctuations. 
 In order to treat this region explicitly, we must carry out 
the self-consistent calculation using the renormalized Green function. 
 We will carry out the self-consistent calculation in the next section, \S4.
 Although there are some differences, the essential results,
such as the gap structure and so on, do not change even in the self-consistent 
calculation.

\section{Self-Consistent Calculation}

 In this section, we self-consistently calculate 
the single particle self-energy and the TDGL parameters on the basis of 
the formalism described in \S2.
 Furthermore, we calculate the superconducting critical temperature 
suppressed by the effects of the fluctuations, and determine the phase diagram.
 
 To begin with, we explain our method of the numerical calculation 
used in this section. 
 We restrict the $\mbox{\boldmath$k$}$- points 
for which the self-energies are calculated 
to the region close to the Fermi surface. The restricted region is 
shown in Fig.12. 
 As we have mentioned in the previous section, the effects of the 
superconducting fluctuations (resonances) are important 
in the vicinity of the Fermi surface, 
and are not important far from the Fermi surface. 
 Except for $a_{1}$, the TDGL parameters mainly depends on 
the electron states near the Fermi surface. 
 Furthermore, the important $ \mbox{\boldmath$k$} $-points 
for the self-consistent calculation of the self-energy  
$ {\mit{\it \Sigma}}^{{\rm R}} (\mbox{\boldmath$k$},\omega) $ 
exist only in the vicinity of $ \mbox{\boldmath$k$} $, 
because the main contribution is given by the terms with the momentum 
in the vicinity of $ \mbox{\boldmath$q$} = 0 $. 
 Therefore, we have only to consider the self-energy near the Fermi surface. 
Thus, our restriction for the momentum space is justified. 

 \begin{figure}[ht]
%  \figureheight{1.0cm}
   \begin{center}
   \epsfysize=5cm
$$\epsffile{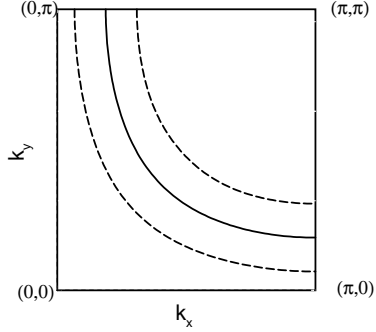}$$
%    \epsfile{file=Fig12,height=5cm}
    \caption{In this section, the self-energy is calculated in the region 
             between the dashed lines. }
     \label{fig:calculated momentum space}
   \end{center}
 \end{figure}

 We estimate the TDGL parameters by eq.~\ref{eq:chi0-general} for the 
renormalized Green function  
${\mit{\it G}}^{{\rm R}} (\mbox{\boldmath$k$},\omega) = 
(\omega - \varepsilon_{\mbox{\boldmath$k$}} - 
{\mit{\it \Sigma}}^{{\rm R}} (\mbox{\boldmath$k$},\omega))^{-1} $, 
and self-consistently calculate the single particle self-energy 
${\mit{\it \Sigma}}^{{\rm R}} (\mbox{\boldmath$k$},\omega) $, 
using eq.~\ref{eq:selfenergyreal} (see Fig.5(b)). 
 Here, we exclude the Hartree-Fock term corresponding to the diagram 
in Fig.4(c), as we have done in the lowest order calculation in \S3. 
 This calculation corresponds to the 
self-consistent T-matrix approximation (see Fig.4(b)). 

 For the purpose of the self-consistent calculation, 
we divide the Brillouin zone as shown in Fig.12, and put 
${\mit{\it \Sigma}}^{{\rm R}} (\mbox{\boldmath$k$},\omega) = 0$ outside of the 
dashed lines. 
 As we have described before, this approximation does not change the structure 
near the Fermi surface and is appropriate to discuss the pseudogap phenomena. 

 Furthermore, we introduce the cutoff parameters for the integrations by 
$\Omega$ and $\mbox{\boldmath$q$}$, because the main contribution is obtained 
from the integrations around $ \mbox{\boldmath$q$} = \Omega = 0 $. 
Owing to the cutoff, we might slightly 
underestimate the effects of the superconducting fluctuations. 
 However, the cutoff procedure has no significant effect on our results. 
 Moreover, the TDGL expansion is not so reliable for large 
$ \mbox{\boldmath$q$} $ and $ \Omega $. 

 It should be noticed that in this self-consistent calculation we fix 
the parameter  $t_{0}$ instead of the coupling constant $g$. 
 Although the final self-consistent results are identical, 
the convergence of the solution is remarkably improved by using this method. 
 In particular, we need to adopt this method in order to calculate near the 
superconducting critical point where the superconducting fluctuations 
are strong. 
 The parameter $t_{0}$ represents the distance from 
the superconducting critical point. Therefore, 
if the parameter $t_{0}$ is varied, 
the situation remarkably changes. As a result, the solution 
fluctuates for the $g$-fixed calculations in many cases. 

 Thus, because of the convenience of the numerical calculation, 
not the coupling constant $g$ but the temperature $T$ and the parameter 
$t_{0}$ are fixed in our calculation. The coupling constant $g$ is a quantity 
determined by a result of the self-consistent calculation. 

 We do not positively change the chemical potential so as to conserve 
the particle number. 
 However, we have verified that the particle number is almost conserved 
in this calculation. 

 Strictly speaking, the superconducting transition does not occur 
in the two-dimensional systems. 
 The fact is known as the Marmin-Wagner's theorem.  
 As we have described in \S2, our calculations also reflect the fact. 
 The superconducting critical temperature $T_{{\rm c}}=0$ in our formalism 
because of the logarithmic singularity of the self-energy  
due to the two-dimensional singularity. 
 However, the weak three-dimensionality is sure to remove these singularities 
in the realistic layered systems.~\cite{rf:randeria}  
 Therefore, we phenomenologically introduce the three-dimensionality and 
define the superconducting critical temperature as the temperature in which 
$ 1 + g \chi_{0}(\mbox{\boldmath$0$},0) = 0.01 $. 
 This condition corresponds to the $100$ times enhancement of the 
superconducting susceptibility. 
 The phase diagram does not depend on the details of this definition, 
qualitatively.

 We show the results of the spectral weight 
for various $t_{0}$ and temperatures in Fig.13.

\begin{figure}[ht]
% \figureheight{1.0cm}
  \begin{center}
   \epsfysize=6cm
$$\epsffile{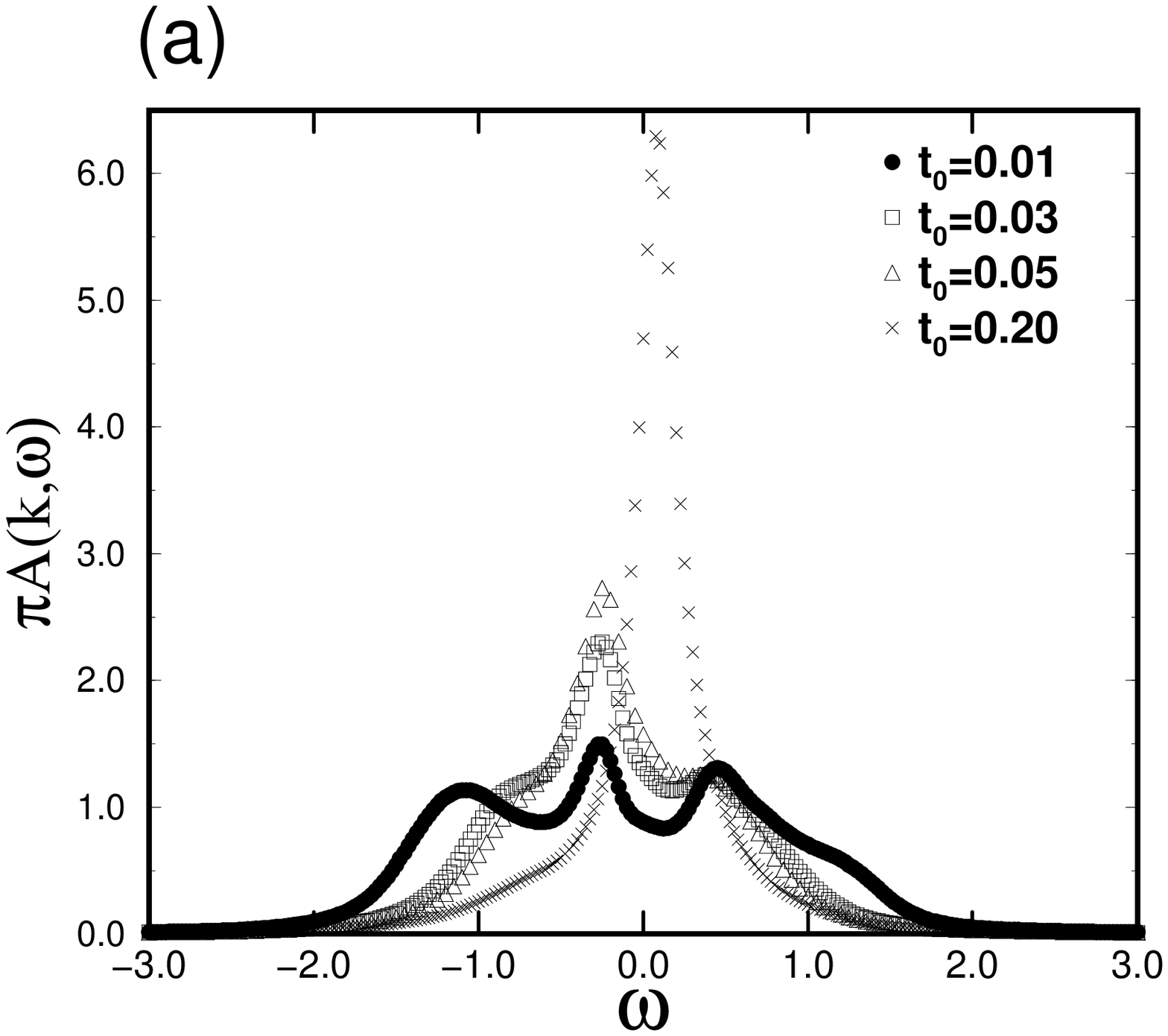}$$
   \epsfysize=6cm
$$\epsffile{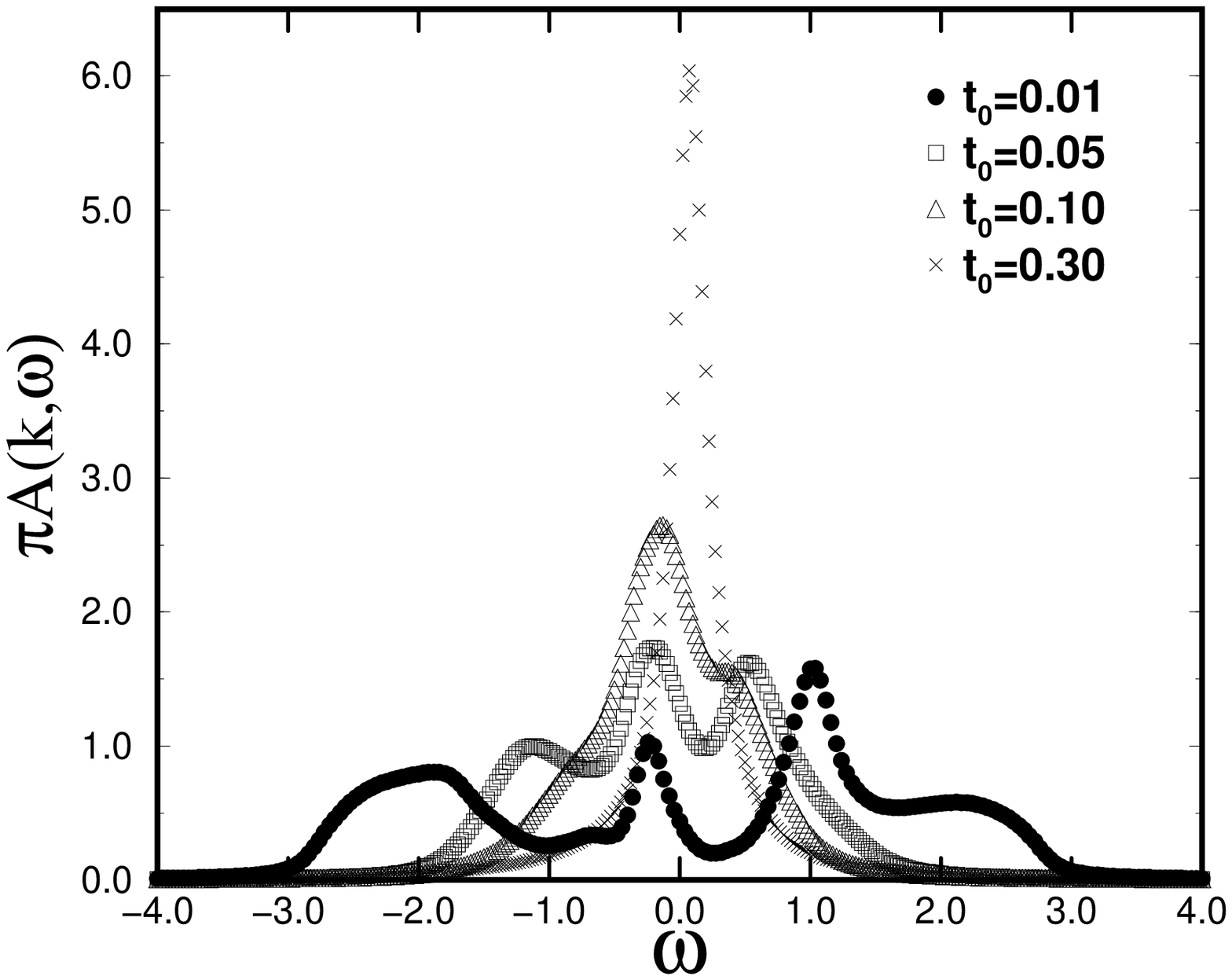}$$
%   \epsfile{file=Fig13a,height=6cm}
%   \epsfile{file=Fig13b,height=6cm}
   \caption{The spectral weight on the Fermi surface 
            near $(0,\pi)$ obtained by the self-consistent calculation. 
            (a)$T=0.10$. $t_{0}$ is varied as $0.01, 0.03, 0.05, 0.20$.
            (b)$T=0.15$. $t_{0}$ is varied as $0.01, 0.05, 0.10, 0.30$. 
            The calculated points are shown in the phase diagram (Fig.17). 
            The $\mbox{\boldmath$k$}$-point is the same as that in Fig.8.  }
    \label{fig:self-consistent-weight}
  \end{center}
\end{figure}

 The gap structure exists in the small $t_{0}$ cases, and becomes clear 
as $t_{0}$ decreases. 
 Then, the properties of the TDGL parameters reproduce our argument in \S2. 
That is to say, 
the superconducting fluctuations are propagative and have 
the small dispersion. 
 Because the parameter $t_{0}$ shows the closeness to 
the superconducting critical point, 
these behaviors are natural.  The effects of the resonances are rather drastic 
in case of the high temperature, that is to say, large $|g|$, 
where the thermal fluctuations are rather strong.

 In particular, the plural peak structure appears in the small $t_{0}$ and 
large $T$ cases. This complicated structure is also understood on the basis of 
the resonance formalism as we explain below.
 For example, we show the self-energy for 
$T=0.10$, $t_{0}=0.01$ in Fig.14.
 In this case, the spectral weight has three peak structure. 

\begin{figure}[ht]
% \figureheight{1.0cm}
  \begin{center}
   \epsfysize=6cm
$$\epsffile{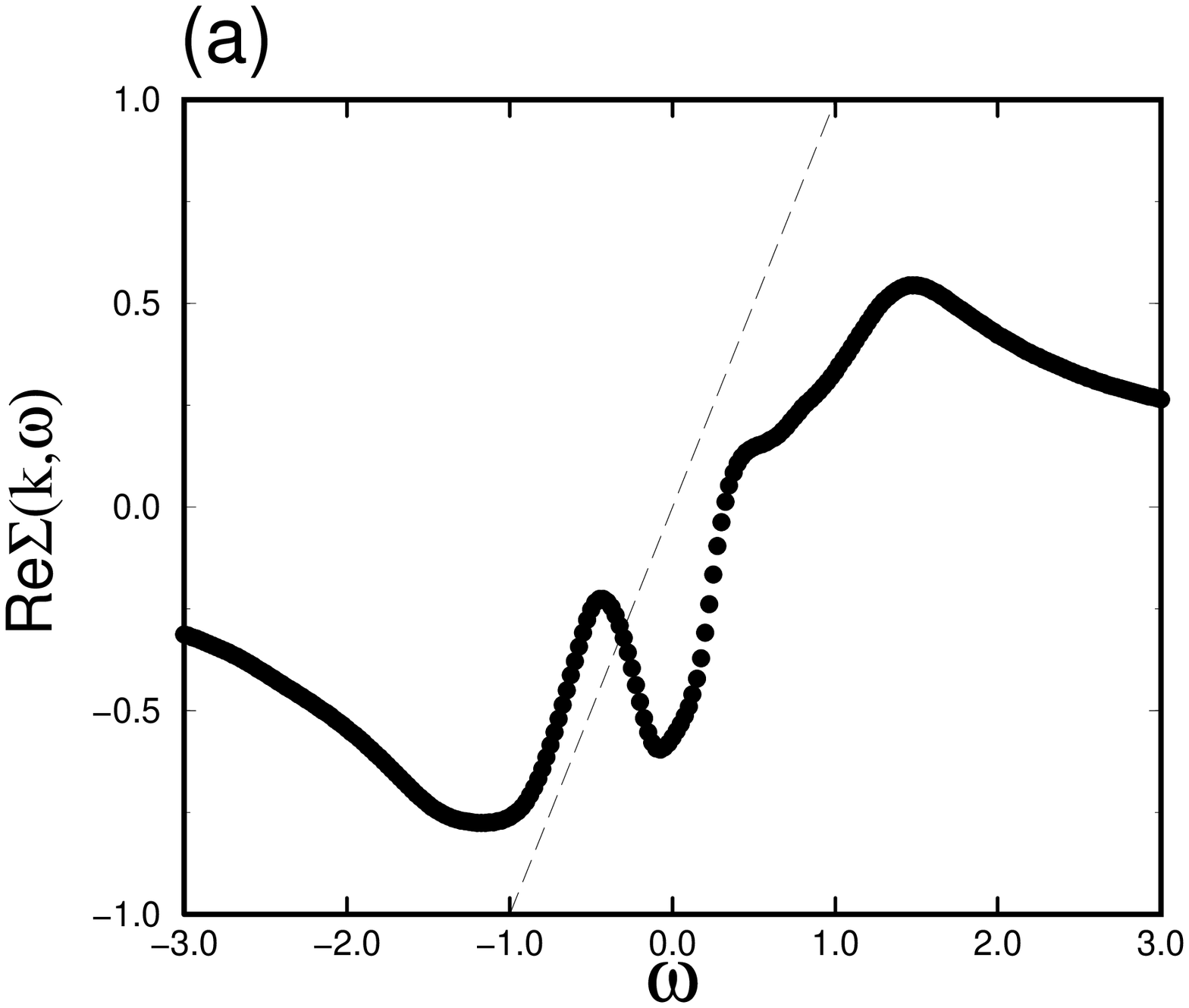}$$
   \epsfysize=6cm
$$\epsffile{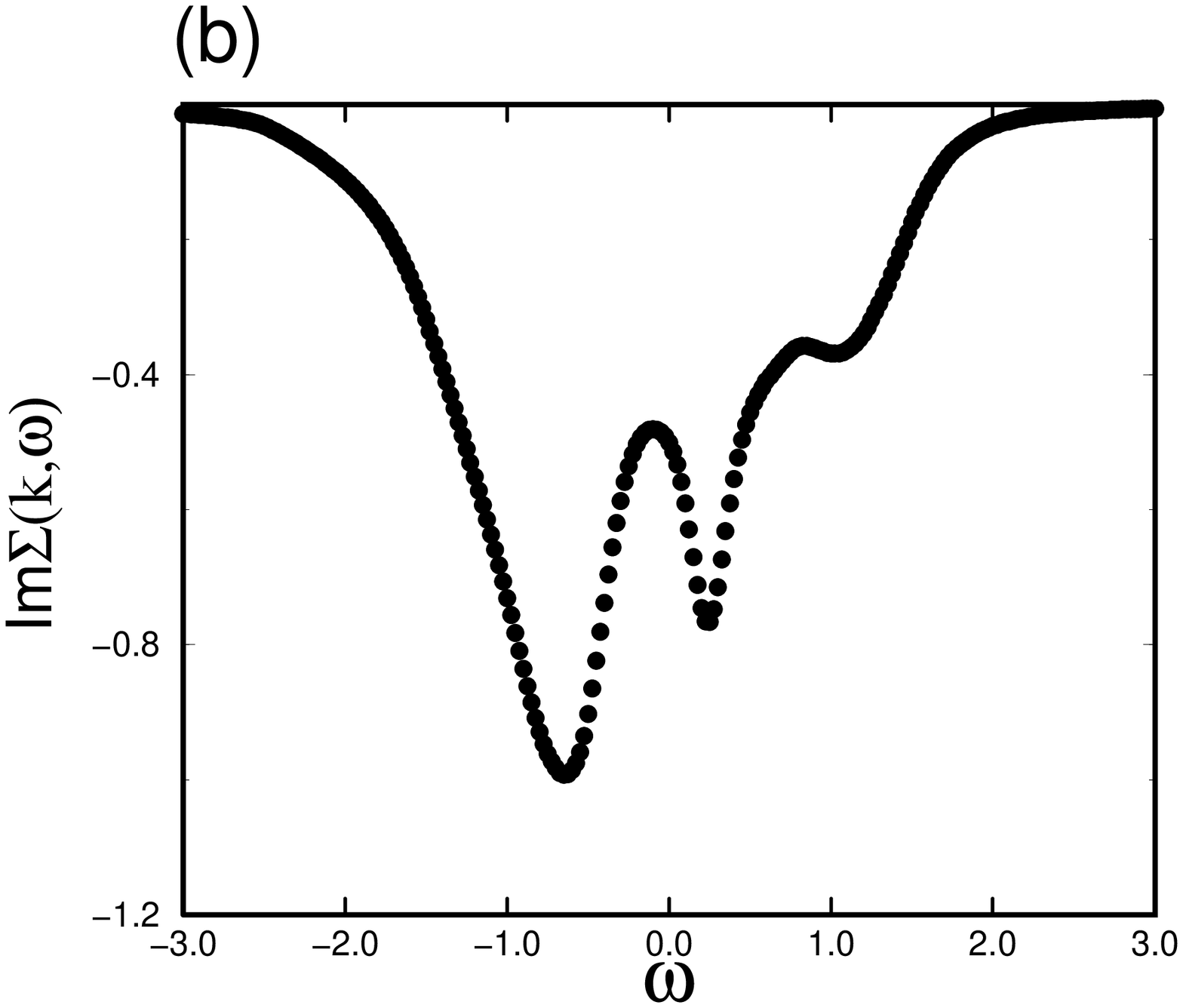}$$
%   \epsfile{file=Fig14a,height=6cm}
%   \epsfile{file=Fig14b,height=6cm}
   \caption{The single particle self-energy on the Fermi surface 
            near $(0,\pi)$. Here, $T=0.10$, $t_{0}=0.01$. 
            (a)The real part. 
            The long-dashed line is described in order to show 
            the line $\omega - \varepsilon_{\mbox{\boldmath$k$}} - 
            {\rm Re}{\mit{\it \Sigma}}^{{\rm R}} (\mbox{\boldmath$k$},\omega) 
            = 0$. 
            (b)The imaginary part. 
            The $\mbox{\boldmath$k$}$-point is the same as that in Fig.8. }
    \label{fig:self-consistent-selfenergy}
  \end{center}
\end{figure}

 It should be noticed that the electronic structure around 
$(-\mbox{\boldmath$k$},-\omega)$ plays an important role for the estimate of 
the self-energy ${\mit{\it \Sigma}}^{{\rm R}} (\mbox{\boldmath$k$},\omega)$.
 It is because the main contribution comes from the integrations 
in the vicinity of $\mbox{\boldmath$q$}=\Omega=0$. 
 Therefore, if the spectral weight has its peak at $\omega=\omega_{{\rm p}}$, 
the real part of the self-energy has the positive slope at 
$\omega=-\omega_{{\rm p}}$, 
and the imaginary part has the peak at 
$\omega=-\omega_{{\rm p}}$ in its absolute value. 
 In rough estimate, 
${\rm Im}{\mit{\it \Sigma}}^{{\rm R}} (\mbox{\boldmath$k$},\omega) 
\propto A(\mbox{\boldmath$k$},-\omega) $. 
 
 To put it in detail, the left peak and the middle peak yield 
the positive slope of the real part on the positive frequency side, 
and yield the right peak. 
 The right peak yields the positive slope of the real part 
on the negative frequency side. 
 Consequently the negative slope is formed at the small negative frequency. 
 The only solution $ \omega - \varepsilon_{\mbox{\boldmath$k$}} - 
{\rm Re} {\mit{\it \Sigma}}^{{\rm R}} (\mbox{\boldmath$k$}, \omega) = 0 $ 
exists here. It corresponds to the middle peak. 
 Moreover, the right peak yields the sharp peak of 
$- {\rm Im}{\mit{\it \Sigma}}^{{\rm R}} (\mbox{\boldmath$k$},\omega)$
at the negative frequency side. The peak yields the double peak structure 
on the negative frequency side. 
 They correspond to the left and middle peaks. 
 It should be noticed that such an asymmetric structure stabilizes 
the self-consistent solution. 
 Because of the rough relation 
${\rm Im}{\mit{\it \Sigma}}^{{\rm R}} (\mbox{\boldmath$k$},\omega) 
\propto A(\mbox{\boldmath$k$},-\omega) $, the peak of the spectral weight 
at $\omega_{{\rm p}}$ will extinguish the peak at $-\omega_{{\rm p}}$. 
 In particular, the spectral weight at the Fermi energy 
$\omega=0$ is necessarily suppressed. 
 Thus, the symmetric structure is unstable in our formalism.
 Therefore, the gap formation is difficult in the particle-hole symmetric 
models, although it is not impossible in the strong coupling case. 

 As we mentioned in \S2, the particle-hole asymmetry naturally exists in the 
real systems and the asymmetry is essential 
for the self-consistent solution to be stabilized. 
 Furthermore, High-$T_{{\rm c}}$ cuprates are 
the strongly asymmetric systems because of existence of 
the Van-Hove singularity. 

 It is notable that these results are not consistent with the assumption by 
Norman {\it et al.}~\cite{rf:norman2} and Maly {\it et al.}~\cite{rf:maly}. 
 Although the rough features of their phenomenological self-energy remain 
in our explicit self-consistent calculation, their assumption cannot 
satisfy the self-consistency.  

 Here, we pay attention to the change of the spectral weight, once more. 
 In the phase diagram (Fig.17), we show the parameters for which 
we show the spectral weight or the density of states. 

 When $t_{0}$ is large and the temperature $T$ is higher than the mean field 
critical temperature $T_{{\rm MF}}$, the spectral weight has 
the sharp single peak structure. 
 This structure is of the conventional Fermi liquid theory. 
 
 As $t_{0}$ decreases and the temperature becomes lower than $T_{{\rm MF}}$, 
the peak shifts to the negative frequency side and 
has the long tail to the positive frequency side. 
 In this region, as the momentum is varied from $(0,\pi)$ to $(\pi,\pi)$, 
the peak shifts to positive frequency side and cross the Fermi level 
$\omega = 0$. 
 However, as a result of resonances, the damping rate 
$- {\rm Im} {\mit{\it \Sigma}}^{{\rm R}} (\mbox{\boldmath$k$}, \omega) $ 
is large at the Fermi level and the spectral weight is suppressed there. 
 Therefore, the density of states is reduced at the Fermi level. 
 The superconducting fluctuations gradually become propagative. 

 As $t_{0}$ decreases further and the system approaches to 
the critical point, the spectral weight has the plural peak structure. 
 This behavior is quite different from that of 
the conventional Fermi liquid theory. 

 Here, we show the density of states for $T=0.10$ and various 
$t_{0}$ in Fig.15.

 \begin{figure}[ht]
   \begin{center}
%    \figureheight{1.0cm} 
   \epsfysize=6cm
$$\epsffile{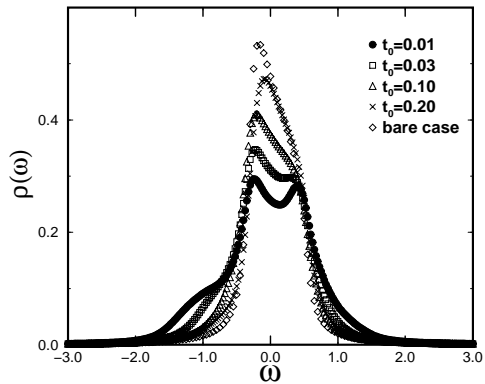}$$
%    \epsfile{file=Fig15,height=6cm}
     \caption{The density of states $ \rho(\omega) $ for $T=0.10$.
              $t_{0}$ is varied as $0.01, 0.03, 0.10, 0.20$.  
              The diamonds show the non-interacting density of states.
              Since we calculate only the states in the vicinity of 
              the Fermi surface, 
              the peak position is not exact. 
              The suppression of the density of states at the low energy 
              is essential.      }
     \label{fig:self-consistent-dos}
   \end{center}
 \end{figure}

 As we mentioned above, the density of states at the Fermi level is reduced 
as a result of the resonance effects. 
 In particular, in case of $t_{0}=0.10$, the spectral weight has 
a sharp single peak in the vicinity of the Fermi surface. However, 
we can see clearly the suppression of the density of states. 
 Generally speaking, the suppression becomes distinguished 
at the mean field critical temperature $T_{{\rm MF}}$. 
 The suppression becomes more remarkable, as $t_{0}$ decreases and the system 
approaches the critical point. 
 In the vicinity of the critical point, the density of states at 
the Fermi level is mainly given by the contribution 
from the quasiparticles near $(\pi/2,\pi/2)$. 
 Therefore, $\rho_{{\rm d}}(0)$ is more remarkably reduced 
by the effects of the resonances.

 Here, we show the momentum dependence of the spectral weight for the typical 
three peak case in Fig.16.

 \begin{figure}[ht]
%  \figureheight{1.0cm}
   \begin{center}
   \epsfysize=6cm
$$\epsffile{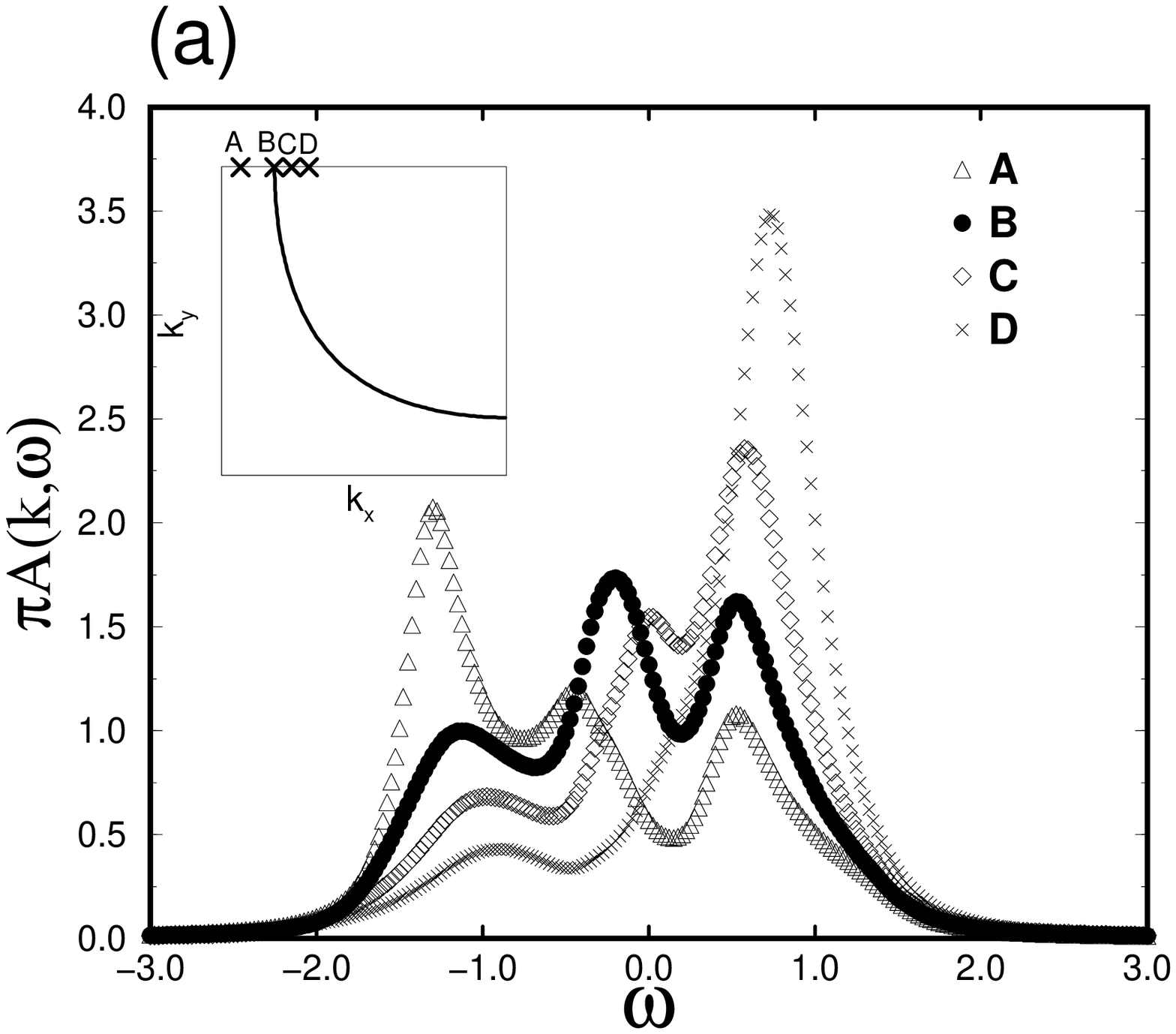}$$
   \epsfysize=6cm
$$\epsffile{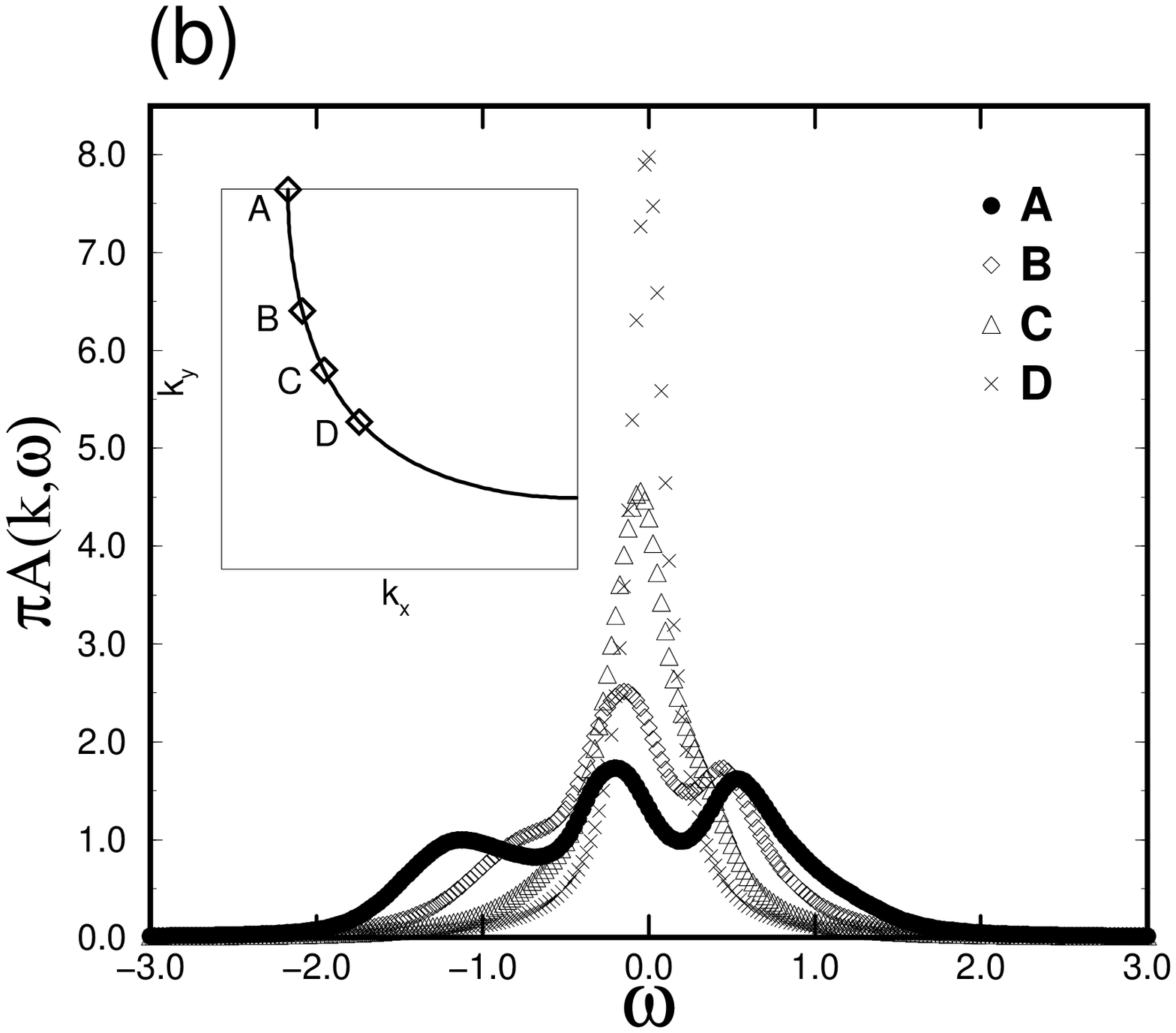}$$
%    \epsfile{file=Fig16a,height=6cm}
%    \epsfile{file=Fig16b,height=6cm}
    \caption{The momentum dependence of the spectral weight. 
             Here, $T=0.15$, $t_{0}=0.05$. The momentum is varied
             (a)across the Fermi surface near $(0,\pi)$ 
             (b)along the Fermi surface. 
             The respective $\mbox{\boldmath$k$}$-points
             are shown in the inset. 
             }
     \label{fig:self-consistent-k}
   \end{center}
 \end{figure}

 In Fig.16(a), the momentum is varied across the Fermi surface near $(0,\pi)$.
 On the negative energy side from the Fermi surface, 
the spectral weight transfers to the left peak. 
 On the other hand, it transfers to the right peak on the opposite side 
from the Fermi surface. 
 All peaks shift to the positive frequency side with the momentum shift. 
 It should be noticed that the middle peak shifts and is found 
at the Fermi level $ \omega = 0 $ 
when the momentum slightly deviates from the Fermi surface 
(represented as C in the inset). However, it is strongly reduced 
by the effects of the fluctuations. 
 With increasing $k_{x}$, this peak cannot cross the Fermi level 
$ \omega = 0 $ 
and disappears at last. Thus, the spectral weight shows the double peak 
structure at D (see the inset). 

 In Fig.16(b), the momentum is varied along the Fermi surface. 
 We can see that the gap structure is remarkable in the vicinity of $(0,\pi)$ 
where the attractive interaction is strong, and becomes inconspicuous 
as the momentum approaches to $(\pi/2,\pi/2)$.  
 Furthermore, the single peak of quasiparticles exists in the vicinity of 
$(\pi/2,\pi/2)$. This fact shows that the Fermi liquid like behavior exists 
in this region. 
 Thus, it is naturally led from our formalism that the pseudogap has 
the same shape as the superconducting gap, as suggested 
by the ARPES experiments~\cite{rf:norman}. 

 Last of all, we show the phase diagram in Fig.17.
 We show the mean field critical temperature $T_{{\rm MF}}$ and 
the actual critical temperature $T_{{\rm c}}$ calculated self-consistently. 

 \begin{figure}[ht]
   \begin{center}
%    \figureheight{1.0cm} 
   \epsfysize=6cm
$$\epsffile{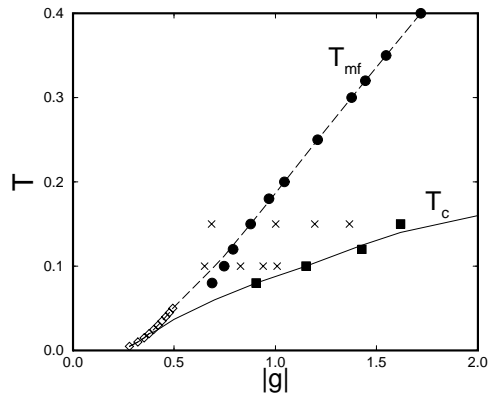}$$
%     \epsfile{file=Fig17,height=6cm}
     \caption{The phase diagram obtained in this paper. 
              The circles and diamonds show the superconducting critical 
              temperature based on the mean field theory ($T_{{\rm MF}}$). 
              The circles are obtained by using the discrete momentum 
              used in our self-consistent calculation. 
              The diamonds are obtained by the explicit calculation 
              for eq.~\ref{eq:chi0}. 
              The circles at the low temperature are slightly off by the 
              numerical error.
              The squares show the actual superconducting critical 
              temperature ($T_{{\rm c}}$) suppressed by the fluctuations. 
              The suppression becomes stronger in the strong coupling region. 
              The marks $\times$ show the points where we have shown 
              the spectral weight or the density of states. 
              They correspond to $t_{0}=0.20$, $t_{0}=0.10$, $t_{0}=0.05$, 
              $t_{0}=0.03$ from left to right for $T=0.10$, and 
              $t_{0}=0.30$, $t_{0}=0.10$, $t_{0}=0.05$, $t_{0}=0.03$ 
              for $T=0.15$, respectively. As we mentioned in this article, 
              the squares correspond to $t_{0}=0.01$. 
                                                             }
     \label{fig:phasediagram}
   \end{center}
 \end{figure}

 The superconducting critical temperature ($T_{{\rm c}}$) is strongly reduced 
by the fluctuations. 
 This effect is mainly caused by the suppression of the density of states. 
 As the coupling constant increases, the reduction of the critical temperature 
becomes remarkable. 
 It should be noticed that this effect is different from that of 
the Nozi$\grave{{\rm e}}$res and Schmitt-Rink formalism. 
 According to the Nozi$\grave{{\rm e}}$res and Schmitt-Rink formalism, 
the critical temperature is reduced by the shift of the chemical potential. 
The chemical potential shifts so that the density of state decreases. 
 In our formalism, the density of states is reduced by the effects of 
the resonances with the thermally excited weakly-damped pre-formed pairs. 

 Roughly speaking, the pseudo-gap phenomena occurs in the region 
between $T_{{\rm MF}}$ and $T_{{\rm c}}$. This region is extremely wide 
in the strong coupling case. 

 The density of states starts to decrease, roughly at $T_{{\rm MF}}$. 
 Most of the physical quantities measured by the various experiments, such as 
NMR,~\cite{rf:NMR,rf:isida} optical conductivity,~\cite{rf:homes} 
tunneling spectroscopy~\cite{rf:renner} and so on, reflect 
the density of state. Therefore, their experiments should show some changes at
$T_{{\rm MF}}$. 
 Actually, various changes around $T_{{\rm MF}}$ are reported 
by several experiments.~\cite{rf:oda} 

 Moreover, the ARPES experiments directly reflect 
the spectral weight. In our calculation, the quasiparticle's peak starts to 
shift to the negative frequency side and become broad at $T_{{\rm MF}}$. 
 As the temperature approaches to $T_{{\rm c}}$, the spectral weight 
at $\omega=0$ is strongly reduced. 
 This reduction of the spectral weight has been observed by the ARPES 
experiments.~\cite{rf:norman} 
 The plural peak structure of the spectral weight is also observed 
by the ARPES experiments.~\cite{rf:takahashi} 
 However, since the spectral weight is extremely broad 
when the plural peak structure appears, 
such a structure may not be observed actually. 
 Moreover, by summing up the momentum around the measured momentum, 
corresponding to the experimental resolving power, 
the plural peak structure may disappear. 
 The reduction of the spectral weight at the Fermi energy $\omega=0$ 
is essential. 
 In this sense, our calculation is consistent with the ARPES experiments. 

 On the other hand, the specific heat starts to decrease 
at rather higher temperature~\cite{rf:loram}. 
 We think that in the strongly correlated electron systems, the entropy mainly 
originates from the freedom of spins. 
 Therefore, the specific heat is expected to 
be sensitive to the anti-ferromagnetic spin fluctuations. 
 The anti-ferromagnetic correlation exists at rather higher temperature 
than $T_{{\rm MF}}$ in high-$T_{{\rm c}}$ cuprates. 
 Although we do not take account of its effects in this paper, 
we consider that the specific heat is reduced 
by the anti-ferromagnetic spin fluctuations. 
 The recent experiment~\cite{rf:ido} shows that the specific heat  
decreases still more, but only slightly at $T_{{\rm MF}}$. 
 This fact is naturally explained by our calculation in this paper 
which shows the suppression of the density of states.

\section{Summary and Discussion}

 In this paper, we have calculated the effects of the strong coupling 
superconductivity on the normal state electronic structure. 
 We have considered that the strong coupling superconductivity takes place 
in under-doped High-$T_{{\rm c}}$ cuprates and the pseudogap phenomena are 
its precursor. 

 We took account of the realistic situation, such as the two-dimensionality, 
the $d$-wave symmetry of the pairs, the lattice system near the half-filling,  
the particle-hole asymmetry and so on. 

 First, we considered the T-matrix which is the dominant scattering vertex 
in the vicinity of the superconducting critical point. 
 Since the T-matrix diverges at $\mbox{\boldmath$q$}=\Omega=0$ 
at the critical point (Thouless criterion), 
we expand the reciprocal of the T-matrix 
with respect to $\mbox{\boldmath$q$}^{2}$ and $\Omega$ (TDGL expansion) 
and study its contribution to the single particle self-energy 
from the small $\mbox{\boldmath$q$}$ and $\Omega$ region. 
 
 We showed that although the superconducting fluctuations are strongly 
diffusive in the weak coupling limit, they are propagative and possess the 
character as the pre-formed pairs in case of the strong coupling 
superconductivity. 
 In particular, the pre-formed pairs have the hole-like character 
reflecting the band structure of High-$T_{{\rm c}}$ cuprates. 

 On the basis of the above argument, we analytically calculated 
the self-energy and showed the anomalous behaviors of the self-energy, 
using the phenomenologically introduced TDGL parameters. 
 In particular, the real part of the self-energy has the large positive slope 
near $\alpha=\omega+\varepsilon_{\mbox{\boldmath$k$}}=0$, 
and the imaginary part has the sharp peak in its absolute value there. 
 These features are quite different from those given by the conventional 
Fermi liquid theory. 
 Such a self-energy leads to the gap-like structure of the spectral 
weight. Furthermore, because of the hole-like character of 
the pre-formed pairs, the asymmetric structure is naturally introduced. 
 The peak of the spectral weight is relatively broad 
on the negative frequency side, and sharp on the positive frequency side. 
 These are the effects of the resonances of quasiparticles with the 
thermally excited weakly-damped pre-formed pairs. 
 
 Furthermore, we explicitly estimated the TDGL parameters 
and calculated the self-energy for the bare Green function 
$ {\mit{\it G}}^{{\rm R(0)}} (\mbox{\boldmath$k$},\omega) $, 
and the renormalized Green function  
$ {\mit{\it G}}^{{\rm R}} (\mbox{\boldmath$k$},\omega) $ which we refer to 
the lowest order calculation and the self-consistent calculation, 
respectively. 
 They correspond to the T-matrix approximation and 
the self-consistent T-matrix approximation, respectively.

 In the lowest order calculation, since the damping rate of 
the pre-formed pairs is not suppressed by the renormalization, 
the effects of the resonances are smaller than those obtained by 
our analytic calculation. 
 However, the self-energy has the same features as those of 
our analytic calculation. 
 As a result, the spectral weight shows the gap-like structure. 

 By carrying out the self-consistent calculation, 
we confirmed the change of the TDGL parameters by the renormalization effects 
which we argued in \S2. 
 Furthermore, the self-energy has the relevant features to the idea 
of the resonances. They are not consistent with the assumption by 
Norman {\it et al.}~\cite{rf:norman2} and Maly {\it et al.}~\cite{rf:maly}. 
 However, the spectral weight necessarily shows the gap-like structure 
in the vicinity of the superconducting critical point, 
although the structure is quite different from that assumed by 
Norman {\it et al.}~\cite{rf:norman2} and Maly {\it et al.}~\cite{rf:maly}. 
 The structure is extremely asymmetric and the spectral weight is strongly 
reduced at the Fermi level. 
 This asymmetry, which is essential for the idea of the resonances,  
stabilizes the self-consistent solution. 
 Since the asymmetric structure is neglected in the assumption by 
Norman {\it et al.}~\cite{rf:norman2} and Maly {\it et al.}~\cite{rf:maly}, 
their assumption cannot satisfy the self-consistency. 
 By summing up the momentum, we obtained the density of states in which  
the complicated structure disappears. As a result, the density of states is 
also suppressed at the Fermi level by the effects of the resonances. 

 The effects of the superconducting fluctuations manifest at the mean field 
superconducting critical temperature $T_{{\rm MF}}$. 
 The density of states starts to decrease at $T_{{\rm MF}}$ and 
the characteristic change appears in the various quantities.  
 These are the pseudogap phenomena and are consistent with 
experiments~\cite{rf:oda}. 

 We obtained the phase diagram by the self-consistent calculation. 
 The superconducting critical temperature $T_{{\rm c}}$ is reduced 
by the suppression of the density of states. 
 The reduction becomes more remarkable as the coupling constant increases. 
 Therefore, although the mean field critical temperature $T_{{\rm MF}}$ 
remarkably increases with the coupling constant $|g|$, $T_{{\rm c}}$
does not vary so much. 
 The pseudogap phenomena take place between $T_{{\rm MF}}$ and $T_{{\rm c}}$. 
 Therefore, the pseudogap exists in the wide region 
in the strong coupling case. 

 Considering that the band width $\varepsilon_{{\rm F}}$ is renormalized 
by the electron-electron correlation, the ratio 
$T_{{\rm c}}/\varepsilon_{{\rm F}}$ is increased by the renormalization.  
 As the doping quantity decreases, the system approaches to 
the Mott insulator. Therefore, it is natural to consider that 
the renormalization effects are enhanced with decreasing doping quantity. 
 Since the anti-ferromagnetic spin fluctuations are enhanced at the same time, 
the attractive interaction becomes strong in the under-doped region. 
 Therefore, it is natural to consider that the superconductivity 
effectively becomes the strong coupling, with decreasing doping quantity. 

 Both $T_{{\rm c}}$ and $T_{{\rm MF}}$ are 
scaled by the band width $\varepsilon_{{\rm F}}$. 
 By considering these facts, it is naturally understood that 
$T_{{\rm c}}$ decreases with the doping rate, although $T^{*}$ 
doesn't so. 
 Since $T_{{\rm c}}/\varepsilon_{{\rm F}}$ is almost independent of 
the coupling constant $|g|$ in the strong coupling region, 
$T_{{\rm c}}$ decreases with $\varepsilon_{{\rm F}}$ in the under-doped 
region. 
 On the other hand, $T_{{\rm MF}}/\varepsilon_{{\rm F}}$ increases with $|g|$. 
 Therefore, $T_{{\rm MF}}$ increases with the attractive interaction $|g|$ 
in spite the decrease of the effective band width 
$\varepsilon_{{\rm F}}$ in the under-doped region. 

 Thus, our theory naturally and appropriately explains 
the pseudogap phenomena in High-$T_{{\rm c}}$ cuprates. 

 Of course, the anti-ferromagnetic spin fluctuations are important 
to explain some experiments. In particular, they play a dominant role 
in the temperature region higher than $T^{*}$. 
 For example, the NMR spin lattice relaxation 
rate $1/T_{1}$ is enhanced at the higher temperature than $T^{*}$ and 
the electronic specific heat is reduced well above $T^{*}$. 
 We consider that these behaviors are attributed to the effects of 
the anti-ferromagnetic spin fluctuations. We do not take account of 
the effects. 
 However, the so-called pseudogap phenomena which are experimentally 
observed bellow $T^{*}$ are naturally understood as the precursor of 
the $d$-wave strong coupling superconductivity. 

 Here, we mention some important factors in our theory. 
 Of course, the strong coupling superconductivity is most essential. 
 The coupling constant $g$ does not appear explicitly. 
 However, it is included in the properties of the TDGL parameters. 
In particular, it is essential that $b$ and $a_{2}$ are reduced 
in the strong coupling case by the renormalization. 
 The lattice system near the half-filling is important for the resonance 
effects. The effects of the resonances are different from that of 
Nozi$\grave{{\rm e}}$res and Schmitt-Rink formalism which is justified in the 
low density limit. 
 The two-dimensionality is essential because it leads to the strong 
fluctuations which are characteristics of the low-dimensional systems. 
 The particle-hole asymmetry is also important to stabilize the 
self-consistent solution. 

 These factors properly reflect the characteristics of 
High-$T_{{\rm c}}$ cuprates. 
 Thus, the realistic treatment carried out in this paper is necessary for the 
pseudogap phenomena. In other words, the pseudogap phenomena well represent 
the individuality of High-$T_{{\rm c}}$ cuprates. 

 Here, it should be noted that although we have emphasized 
the importance of the particle-hole asymmetry, 
the pseudogap formation is possible 
in principle even in the particle-hole symmetric case. 
(Actually, this is a perfect nesting case and is not realistic.) 
 In the symmetric case $a_{1}=0$, and the superconducting fluctuations are 
diffusive. However, it is the same that the small 
$\mbox{\boldmath$q$}$ and $\Omega$ region make a dominant contribution for the 
self-energy. 
 Therefore, the strong fluctuations necessarily make the Fermi liquid 
state unstable even in the symmetric models. 
 After all, the pseudogap formation is relatively difficult but not 
impossible in the symmetric case. 

 Here, we give a discussion on the magnetic scenarios for the pseudogap 
phenomena. The calculation treating the pseudogap phenomena as a precursor of 
the magnetic instability (anti-ferromagnetism or SDW) has been 
calculated.~\cite{rf:dahmSDW} 
 We also think it should be possible that the strong spin fluctuations 
near the magnetic critical point lead to the gap-like structure. 
 In that case, the gap formation first takes place at 'hot spot', because 
the interaction is strong in the vicinity of $\mbox{\boldmath$q$}=(\pi,\pi)$ 
and $\Omega=0$. 
 Since the spin fluctuations lead to the transformation of the Fermi surface, 
'hot spot' comes to exist in the wide region near $(0,\pi)$.~\cite{rf:yanase}  
 Therefore, the gap structure would have a similar shape to 
the $d_{x^{2}-y^{2}}$-wave superconducting gap. 

 However, we think it does not occur in High-$T_{{\rm c}}$ cuprates. 
 Since the interaction caused by the superconducting fluctuations 
is strong in the vicinity of $\mbox{\boldmath$q$}=(0,0)$, 
the important $\mbox{\boldmath$k$}$-points for the self-energy 
on the Fermi surface are sure to be on the Fermi surface. 
 On the other hand, in case of the anti-ferromagnetic spin fluctuations, 
the interaction is strong in the vicinity of $\mbox{\boldmath$Q$}=(\pi,\pi)$. 
 Therefore, the important $\mbox{\boldmath$k$}$-points for the quasiparticles 
with the momentum $\mbox{\boldmath$k$}$ exist in the vicinity of 
$\mbox{\boldmath$k$}+\mbox{\boldmath$Q$}$. 
 They are on the Fermi surface only when $\mbox{\boldmath$k$}$ is on the 
'hot spot'.~\cite{rf:yanase} 
 As a result, the quasiparticles slightly apart from 'hot spot' are not 
directly scattered by the strong interaction. 
 Thus, the pseudogap formation is not impossible but difficult to be 
attributed to the magnetic interaction on the numerical point of view.  
 Moreover, since the phase transition which really occurs is 
the superconductivity, the superconducting critical phenomena necessarily 
take place. 
 It is not obvious how the critical spin fluctuations can exist 
even in the critical region of the superconductivity. 
 Indeed, the spin fluctuations have turned out to be reduced in the pseudogap 
region. 
 Furthermore, the continuity at the superconducting critical point is 
difficult to be explained naturally by the magnetic scenarios.  

 On the other hand, a slight suppression of the density of states is indicated 
at the rather higher temperature than $T^{*}$.~\cite{rf:renner} 
 We expect that it is derived from the effects of 
the anti-ferromagnetic spin fluctuations. 
 
 Here, we shortly discuss the magnetic field dependence of the pseudogap 
phenomena. A recent experiment by Gorny {\it et al.}~\cite{rf:gorny} 
has reported that the NMR spin lattice relaxation rate $1/T_{1}$ has 
no magnetic field dependence around the pseudogap onset temperature $T^{*}$. 
 On the other hand, Mitrovi$\acute{{\rm c}}$ {\it et al}. have 
reported that $1/T_{1}$ has the magnetic field dependence~\cite{rf:Mitrovic} 
and given the comment that their sample is optimally-doped and 
that of Gorny {\it et al.} is under-doped.~\cite{rf:Mitroviccomment} 
 Because the pseudogap phenomena continuously take place from optimally-doped 
to under-doped cuprates, these magnetic field dependences also should be 
understood continuously. 

 The pseudogap effects are expected to be relatively small 
in optimally-doped cuprates. 
 Eschrig {\it et al.} have shown that the magnetic field 
dependence of the optimally-doped cuprates is consistent with 
the results of the 
conventional $d$-wave superconducting fluctuation theory.~\cite{rf:eschrig} 
 In their calculation, the DOS corrections play an important role. 
 As we mentioned in the previous sections, 
our theory is a natural extension of their conventional approach. 

 As the coupling constant $|g|$ increases, the characteristic length 
$\xi \propto b^{\frac{1}{2}}$ 
of the orbital motion of the pre-formed pairs decreases. 
 Generally speaking, the main effect of the magnetic field is  
the Landau quantization. Owing to the Landau quantization, 
$\mbox{\boldmath$q$}^{2}$ is quantized. The coefficient of 
$\mbox{\boldmath$q$}^{2}$ is $b \propto \xi^{2}$ and is small in the strong 
coupling case. Moreover, $t_{0}$ is relatively large near the 
onset temperature $T^{*}$. 
 Since the effects of the magnetic fields are actually scaled by $b/t_{0}$, 
the effects are further small near $T^{*}$. 
 Therefore, we think that the magnetic field 
dependence is relatively weak in the strong coupling case, 
that is, in under-doped cuprates. 
 The fact is especially concluded near $T^{*}$. 
 Thus, the pairing scenario for the pseudogap phenomena discussed 
in this paper does not contradict with the continuous comprehension of 
the magnetic field dependences. 

 The continuity of the phase diagram of High-$T_{{\rm c}}$ cuprates are 
understood as described bellow on the basis of our scenario. 
 As the doping rate decreases from the over-doped region, the band width 
$\varepsilon_{{\rm F}}$ is renormalized and the critical temperature 
$T_{{\rm c}}$ increases. 
 Then, $T_{{\rm c}}/\varepsilon_{{\rm F}}$ gradually increases, and 
the effects of the superconducting fluctuations appear. 
 When the doping rate decreases further and 
cross the optimally-doped region, the strong coupling superconductivity 
described in this paper is realized and the pseudogap phenomena occur  
in the wide temperature region. 

 The definite calculations for the physical quantities such as 
the spin lattice relaxation rate, the spin susceptibility, the specific heat, 
the transport coefficients and so on are the important future problems. 

 As we have mentioned above, the strong coupling superconductivity takes place 
in the electron systems renormalized by the strong repulsive interaction. 
 The formulation of the strong coupling superconductivity for the 
strongly renormalized quasiparticles is a greatly important future problem.

\section*{Acknowledgements}

 The authors are grateful to Professor M. Ido, Professor M. Oda and 
Dr. N. Momono for fruitful discussions, 
and to Professor T. Takahashi for valuable comments. 
 Numerical computation in this work was partly carried out 
at the Yukawa Institute Computer Facility. 
 The present work was partly supported by a Grant-In-Aid for Scientific 
Research from the Ministry of Education, Science, Sports and Culture, Japan. 
 One of the authors (Y.Y) has been supported by a Research Fellowships of the 
Japan Society for the Promotion of Science for Young Scientists.

\end{document}